\documentclass[aps,twocolumn,showpacs,pra,superscriptaddress,amsmath,amssymb]{revtex4-1}
\newcommand{\bras}[1]{\langle#1|}
\newcommand{\kets}[1]{|#1\rangle}
\newcommand{\brasd}[1]{\langle\!\langle#1|}
\newcommand{\ketsd}[1]{|#1\rangle\!\rangle}
\newcommand{\bra}[1]{\left<#1\right|}

\newcommand{\means}[1]{\langle#1\rangle}

\usepackage{graphicx}
\usepackage{dcolumn}
\usepackage{bm}
\usepackage{amsmath}
\usepackage[usenames]{color}

\usepackage{txfonts}
\usepackage[T1]{fontenc}
\usepackage{xspace}

\usepackage{hyperref}

\usepackage{ulem}

\begin{document}
%\draft
%\preprint{?????}
\title{
Phase diagram and collective excitation in excitonic insulator: from orbital physics viewpoint
}
\author{Joji Nasu}
\affiliation{
  Department of Physics, Tokyo Institute of Technology, 
  Meguro, Tokyo 152-8551, Japan
} 
\author{Tsutomu Watanabe}
\affiliation{
  Department of Natural Science, Chiba Institute of Technology,
  Narashino, Chiba 275-0023, Japan
} 
\author{Makoto Naka}
\affiliation{
  Department of Physics, Tohoku University, Sendai 980-8578, Japan
} 
\author{Sumio Ishihara}
\affiliation{
  Department of Physics, Tohoku University, Sendai 980-8578, Japan
} 

 \date{\today}
 \begin{abstract}
An excitonic-insulating system is studied from a viewpoint of the orbital physics in strongly correlated electron systems.  
An effective model Hamiltonian for low-energy electronic states is derived from the two-orbital Hubbard model with a finite energy difference corresponding to the crystalline-field splitting. 
The effective model is represented by the spin operators and the pseudo-spin operators for the spin-state degrees of freedom.
The ground state phase diagram is analyzed by the mean-field approximation. 
In addition to the low-spin state and high-spin state phases, two kinds of the excitonic-insulating phases emerge as a consequence of the competition between the crystalline-field effect and the Hund coupling. 
Transitions to the excitonic phases are classified to an Ising-like transition resulted from a spontaneous breaking of the $Z_2$ symmetry.
Magnetic structures in the two excitonic-insulating phases are different from each other; 
an antiferromagnetic order and a spin nematic order. 
Collective excitations in each phase are examined using the generalized spin-wave approximation. 
Characteristics in the Goldstone modes in the excitonic-insulating phases are studied through the calculations of the dynamical correlation functions for the spins and pseudo-spin operators.
Both the transverse and longitudinal spin excitation modes are active in the two excitonic-insulating phases in contrast to the low-spin state and high-spin state phases. 
Relationships of the present results to the perovskite cobalt oxides are discussed. 
\end{abstract}

%\pacs{75.25.Dk,75.70.Tj,75.10.Jm,75.30.Et}
%75.10.Kt	Quantum spin liquids, valence bond phases and related phenomena
%75.10.Jm	Quantized spin models, including quantum spin frustration
%03.67.Pp	Quantum error correction and other methods for protection against decoherence
%71.10.Pm	Fermions in reduced dimensions (anyons, composite fermions, Luttinger liquid, etc.)
%03.67.Lx	Quantum computation architectures and implementations
%75.30.Et	Exchange and superexchange interactions
%71.70.Ej	Spin-orbit coupling, Zeeman and Stark splitting, Jahn-Teller effect
%71.35.Lk	Collective effects (Bose effects, phase space filling, and excitonic phase transitions)
%73.43.Nq	Quantum phase transitions (see also 64.70.Tg Quantum phase transitions in equations of state, phase equilibria and phase transitions)
%75.30.Kz	Magnetic phase boundaries (including classical and quantum magnetic transitions, metamagnetism, etc.)
%74.25.Bt	Thermodynamic properties

%75.25.Dk 	Orbital, charge, and other orders, including coupling of these orders
%75.30.Et 	Exchange and superexchange interactions (see also 71.70.Gm Exchange interactions)
%75.47.Lx 	Magnetic oxides
%75.70.Tj	Spin-orbit effects (see also 71.70.Ej Spin-orbit coupling, Zeeman and Stark splitting, Jahn-Teller effect)

\maketitle

\section{Introduction}
%%%%%

%%%%%
%% excitonic insulator

Excitonic condensation and excitonic-insulating (EI) state have been studied for a long time since the theoretical proposal in 1960s~\cite{mott1961transition,PhysRev.158.462,RevModPhys.40.755,PhysRevB.62.2346,PhysRevLett.89.166403,0953-8984-27-33-333201,doi:10.1143/JPSJ.31.730,doi:10.1143/JPSJ.31.812,doi:10.1143/JPSJ.44.1759}.  
When the exciton binding energy exceeds the band-gap energy in semiconductors and semimetals, a macroscopic number of excitions exhibit a condensation. 
This phenomenon is regarded as a spontaneous quantum-mechanical mixing of the conduction band and the valence band, in which one-body electron hoppings are prohibited due to the orbital and lattice symmetries.
Because of the spontaneous mixing between the electron and hole wave functions, analogue to the superconductivity and the charge density wave (CDW) state have been examined. 
Although much efforts have been devoted to search the EI materials, e.g., the group IV and V materials, the compound semiconductors, the rare-earth compounds, and so on, clear evidence to prove the EI phase has been hardly provided experimentally so far. 
Recent intensive studies in the layered chalcogenide Ta$_2$NiSe$_5$ show that this material is a strong candidate for the EI phase~\cite{PhysRevLett.103.026402,Wakisaka2012,PhysRevB.87.035121,PhysRevB.90.245144}.
These studies trigger reinvestigations of the EI state from the modern theoretical and experimental viewpoints in several materials, for example, 1$T$-TiSe$_2$~\cite{PhysRevLett.105.176401,PhysRevB.91.205135,porer2014non}, iron pnictides~\cite{0295-5075-86-1-17006,doi:10.1143/JPSJS.77SC.158,PhysRevB.79.180504,PhysRevLett.110.207205} and others. 

%%%%%
%% cobaltites

The perovskite cobalt oxides are another candidate materials for the EI state.
A series of the cobalt oxides are the well-known system, in which the spin-state transition in a cobalt ion is exhibited. 
Three spin states are possible in a trivalent cobalt ion due to the competition between the crystalline-field effect and the Hund coupling:
the high-spin (HS) state of the $(e_g)^2 (t_{2g})^4$ electron configuration with the spin magnitude $S=2$, 
the intermediate-spin (IS) state of $(e_g)^1 (t_{2g})^5$ with $S=1$, and the low-spin (LS) state of $(t_{2g})^6$ with $S=0$.
The spin state changes from LS to HS or IS are supposed to be realized in LaCoO$_3$ with increasing temperature and by the chemical hole doping and the photo carrier doping~\cite{PhysRevB.58.R1699}.
A possibility of the EI phase is recently suggested in Pr$_{0.5}$Ca$_{0.5}$CoO$_3$~\cite{PhysRevB.66.052418,doi:10.1143/JPSJ.73.1987,hejtmanek2013phase}.
A phase transition occurs around 90~K below which the electrical resistivity increases and the magnetization is reduced.
No superlattice diffraction peaks attributable to magnetic and charge orders have been observed until now. 
It is expected that a nominal valence of the Co ions is expected to be trivalent and the spin-state degree of freedom remains~\cite{hejtmanek2013phase}.
The first principle calculation combined with the dynamical mean-filed theory as well as the model calculations based on the two-orbital Hubbard model suggest a possibility of the EI phase in this material~\cite{PhysRevB.89.115134,PhysRevB.90.235112}. 

%% orbital degree of freedom

From a viewpoint of the strongly correlated electron systems, this issue of the EI phase can be regarded as the physics of the orbital degree of freedom and the electronic multipoles.
It is certain that a finite energy-gap opens between the conduction and valence bands, and the orbital degeneracy seems to be irrelevant in the EI systems.
In spite of the fact, the orbital physics provides an appropriate viewpoint to capture the essence of the EI systems.
The EI transition can be identified as spontaneous off-diagonal orders between the orbitals which compose the valence and conduction bands.
The orthorhombic-monoclinic structural phase transition and the x-ray diffuse scattering observed in Ta$_2$NiSe$_5$ are interpreted as a consequence of the orbital order accompanied by a ferroelastic transition.
The perovskite cobaltites are suitable systems, to which the concept of the correlated electron orbitals are applicable. 
There is a great number of studies in the orbital degrees of freedom under strong electron correlation in $3d$ transition-metal oxides~\cite{Tokura462}. 
Not only the characteristics in the orbital orders and excitations~\cite{PhysRevLett.81.582,PhysRevLett.88.106403,saitoh2001observation,PhysRevB.67.045108}, but also entanglements with other degrees of freedom provide a rich variety of electric, magnetic and optical phenomena~\cite{PhysRevLett.78.2799,PhysRevB.88.094408}. 
Now, the EI phase and the EI phase transition should be reexamined in a framework of the orbital physics.

In this paper, we study the ground-state phase diagram and the collective excitations in the EI system from the orbital physics viewpoint under the strong electron correlation. 
We analyze the effective model Hamiltonian derived from the two-orbital Hubbard model with a finite energy difference between the orbitals.  
Instead of the three spin states in a Co$^{3+}$ ion with the five $3d$ orbitals, we reduce these orbitals to the two orbitals with a finite energy gap at each lattice site.
The two kinds of the spin states termed the LS and HS states are realized as a result of the competition between the Hund coupling and the energy difference between the orbitals. 
The effective model Hamiltonian derived from the two-orbital Hubbard model is represented by the spin operators and the pseudo-spin (PS) operators which describe the spin-state degrees of freedom. 
This model is appropriate to study the symmetries in the ordered phases and the phase transitions, and the low-energy excitations as well as the finite-temperature electronic states. 
The phase diagram at zero temperature is calculated by the mean-field (MF) approximation. 
Two kinds of the EI phases appear between the LS and HS phases. 
Magnetic structures are distinguished in the two EI phases: an antiferromagnetic order and a spin nematic order. 
The collective excitations in each phase are analyzed using the generalized spin-wave theory.
The Goldstone modes emerge in the two EI phases. 
Not only the transverse magnetic modes but also the longitudinal magnetic modes are active in the EI phases. 
All numerical results are consistent with the symmetry analyses.

%%%%%
%% In organization

In Sec.~\ref{sec:model}, the effective model Hamiltonian is derived from the two-orbital Hubbard model. 
In Sec.~\ref{sec:method}, the MF approximation and the generalized spin-wave theory are presented. 
The numerical results for the ground-state phase diagram and the excitation spectra are shown in Sec.~\ref{sec:ground-state-phase} and Sec.~\ref{sec:coll-excit}, respectively.
Section~\ref{sec:discussion-summary} is devoted to discussion and summary.

 \section{Model}\label{sec:model}

We start from a tight-binding model, where two orbitals with a finite energy difference are located at each site. 
The electron hopping integrals between the nearest-neighbor (NN) sites are assumed to be finite between the same kinds of the orbitals.
The on-site electron-electron interactions are taken into account.
 The two-orbital Hubbard model is defined by
\begin{align}
 {\cal H}={\cal H}_t+{\cal H}_U. 
\label{eq:exH}
\end{align}
The first term represents the electron transfer between the NN sites given by 
\begin{align}
 {\cal H}_t=-\sum_{\means{ij}\eta\sigma}t_\eta (c_{i\eta \sigma}^\dagger c_{j\eta \sigma}+{\rm H.c.}), \label{eq:6}
\end{align}
where $c_{i\eta \sigma}^\dagger$ ($c_{i\eta \sigma}$) is the creation (annihilation) operator for an electron with orbital $\eta (=a,b)$ and spin $\sigma (=\uparrow,\downarrow)$ at site $i$, and $t_\eta$ is the transfer integral between the NN sites $i$ and $j$ with the orbital $\eta$. 
The second term in Eq.~(\ref{eq:exH}) represents the on-site energy and the electron-electron interactions given by 
\begin{align}
 {\cal H}_U&=\Delta\sum_i n_{i a} +U\sum_{i \eta}n_{i\eta\uparrow}n_{i\eta\downarrow}+U'\sum_{i} n_{ia}n_{ib}\nonumber\\
&+J\sum_{i\sigma\sigma'}c_{ia\sigma}^\dagger c_{ib\sigma'}^\dagger c_{ia\sigma'}c_{ib\sigma}
 +I\sum_{i\eta\neq \eta'}c_{i\eta\uparrow}^\dagger c_{i\eta \downarrow}^\dagger c_{i\eta'\downarrow}c_{i\eta'\uparrow},\label{eq:3}
\end{align}
where we define the number operators as $n_{i\eta\sigma}=c_{i\eta\sigma}^\dagger c_{i\eta\sigma}$ and $n_{i\eta}=\sum_\sigma n_{i\eta\sigma}$.
The first term in Eq.~(\ref{eq:3}) represents a difference between the energy levels for the orbitals $a$ and $b$, and $\Delta$ is set to be positive. 
The remaining terms are for the on-site electron-electron interactions, where $U$, $U'$, $J$, and $I$ are the intra-orbital Coulomb interaction, the inter-orbital Coulomb interaction, the Hund coupling, and the pair-hopping interaction, respectively. 
The electron-number per site is two on average. 
In an isolated ion, there are the relations $U=U'+2J$ and $I=J$.
The electronic structures in the two-orbital Hubbard model have been investigated so far from the several points of view~\cite{PhysRevB.89.115134,PhysRevB.90.235112,PhysRevLett.99.126405,PhysRevB.80.054410,PhysRevLett.107.167403,PhysRevB.86.045137,PhysRevLett.99.126405,PhysRevB.80.054410,PhysRevB.85.165135}.

In order to examine the low-energy electronic structures, we derive an effective model Hamiltonian from the two-orbital Hubbard model.
The first and second terms in Eq.~(\ref{eq:exH}) are treated as the perturbed and unperturbed terms, respectively, and the electron number is fixed to be two at each site. 
Among the 6[$={4 \choose 2}$] eigenstates in ${\cal H}_U$, we adopt one of double-occupied spin-singlet states and the spin-triplet states as the basis states in the low-energy Hilbert space.
These are termed the LS and HS states, respectively, in the two-orbital Hubbard model, from now on.
This choice of the basis set for the low-energy electronic structure is justified in the case where the crystalline-electric field and the Hund coupling compete with each other. 
The wave function for the spin-singlet state is given by 
\begin{align}
 \kets{L}&=\left(f c_{b\uparrow}^\dagger c_{b\downarrow}^\dagger - g c_{a\uparrow}^\dagger c_{a\downarrow}^\dagger\right)\kets{0}, 
\label{eq:ls}
\end{align}
where $\kets{0}$ is the vacuum for electrons, and the factors are given as  $f=1/\sqrt{1+(\Delta'-\Delta)^2/I^2}$ and $g=\sqrt{1-f^2}$. 
The energy of this state is $E_L=U+\Delta-\Delta'$ with $\Delta'=\sqrt{\Delta^2+I^2}$. 
When $I$ is zero, $f=1$, $g=0$ and $E_L=U$. 
The wave functions for the spin-triplet states are given by 
\begin{align}
 \kets{H_{+1}}&=c_{a\uparrow}^\dagger c_{b\uparrow}^\dagger\kets{0}, 
\label{eq:hs+1} \\
 \kets{H_{0}}&=\frac{1}{\sqrt{2}}\left(c_{a\uparrow}^\dagger c_{b\downarrow}^\dagger + c_{a\downarrow}^\dagger c_{b\uparrow}^\dagger\right)\kets{0},
\label{eq:hs0} \\
 \kets{H_{-1}}&=c_{a\downarrow}^\dagger c_{b\downarrow}^\dagger\kets{0},
\label{eq:hs-1}
\end{align}
with the energy $E_H=\Delta+U'-J$. 
The energy difference between the spin-singlet and spin-triplet states are $E_H-E_L=\Delta'-J+U'-U$, which is $\Delta-J$ in the case of $I=0$.
Thus, the stability of the two spin states is mainly controlled by $\Delta$ and $J$. 
The basis set of the wave functions is denoted as $\{ |H_{+1}\rangle, |H_{0}\rangle, |H_{-1}\rangle,  |L \rangle \}$. 
It is convenient to introduce the equivalent set of the wave functions $\{ |H_X \rangle, |H_Y \rangle, |H_Z \rangle,  |L \rangle \}$,
where we define $|H_X\rangle=(-|H_{+1} \rangle +|H_{-1}\rangle)/\sqrt{2}$, $|H_Y\rangle=-(|H_{+1} \rangle +|H_{-1}\rangle)/(\sqrt{2} i)$, and $|H_Z\rangle=|H_0 \rangle $. 

We present explicit representations of the local operators in the basis set introduced above.  
The spin operators with the magnitude of $S=1$ are represented in the basis set $\{ |H_X \rangle, |H_Y \rangle, |H_Z \rangle,  |L \rangle \}$ as  
\begin{align}
S^x&=\frac{1}{\sqrt{2}}
\left(
\begin{array}{ccc|c}
&&&\\
&&-i&\\
&i&&\\ \hline
&&&
\end{array}
\right), 
\\
 S^y&=\frac{1}{\sqrt{2}}
\left(
\begin{array}{ccc|c}
&&i&\\
&&&\\
-i&&&\\ \hline
&&&
\end{array}
\right), \\
 S^z&=\frac{1}{\sqrt{2}}
\left(
\begin{array}{ccc|c}
&-i&&\\
i&&&\\
&&&\\ \hline
&&&
\end{array}
\right). 
\end{align}
In order to describe the spin-state degree of freedom, we introduce the PS operators $\tau_{\Gamma}^\gamma$ with subscripts $\gamma(=x, y, z)$ and $\Gamma(=X, Y, Z)$. 
The $X$ components of the PS operators, $\tau_X^\gamma$, are represented in the basis set $\{ |H_X \rangle, |H_Y \rangle, |H_Z \rangle,  |L \rangle \}$ as   
\begin{align}
\tau_X^x&=
\left(
\begin{array}{ccc|c}
&&&1\\
&&&\\
&&&\\ \hline
1&&&
\end{array}
\right), 
\\
 \tau_X^y&=
\left(
\begin{array}{ccc|c}
&\ &\ &-i\\
&&&\\
&&&\\ \hline
i&&&
\end{array}
\right), \\
\tau_X^z&=
\left(
\begin{array}{ccc|c}
1&&&\\
&&&\\
&&&\\ \hline
&&&-1
\end{array}
\right) . 
\end{align}
The other components, $\tau_Y^\gamma$ and $\tau_Z^\gamma$, are defined from $\tau_X^\gamma$ by the cyclic permutations of $\{|H_X\rangle, |H_Y\rangle, |H_Z\rangle\}$. 
It is worth to note that $\tau_\Gamma^z$ represents a difference between the weights in the LS and HS states, and $\tau_\Gamma^x$ and $\tau_\Gamma^y$ represent the mixing between the LS and HS states with the real and complex coefficients, respectively.
Therefore, $\means{\tau_\Gamma^x}$ and $\means{\tau_\Gamma^y}$ are the order parameters for the EI phase.
The matrix elements expressed in the basis set $\{ |H_{+1}\rangle, |H_{0}\rangle, |H_{-1}\rangle,  |L \rangle \}$ are obtained by the unitary transformation with the matrix given by 
\begin{align}
U=
\left(
\begin{array}{ccc|c}
\frac{-1}{\sqrt{2}}& \frac{i}{\sqrt{2}}& &\\
&&1&\\
\frac{1}{\sqrt{2}}&\frac{i}{\sqrt{2}}&&\\ \hline
&&&1
\end{array}
\right) .
\end{align}
The PS operators are also represented by the projection operators as 
$\tau_X^{x}=\kets{L}\bra{H_X}+\kets{H_X}\bra{L}$, $\tau_X^{y}=i(\kets{L}\bra{H_X}-\kets{H_X}\bra{L})$,  and $\tau_X^{z}=\kets{H_X}\bra{H_X}-\kets{L}\bra{L}$. 
We introduce the projection operators defined by $n^{\rm H}=\sum_{\Gamma}\kets{H_\Gamma}\bras{H_\Gamma}$ and $n^{\rm L}=\kets{L}\bras{L}$. 
It is useful to introduce another set of the operators 
$d_{+1}=\kets{L}\bra{H_{+1}}$, $d_{0}=\kets{L}\bra{H_{0}}$, and  $d_{-1}=\kets{L}\bra{H_{-1}}$, 
and their combinations 
$d_{X}=(-d_{+1}+d_{-1})/\sqrt{2})$, $d_{Y}=(d_{+1}+d_{-1})/\sqrt{2i})$ and 
$d_{Z}=d_{0}$~\cite{PhysRevB.89.115134}. 
When the electron-hole transformations in the $b$ band are performed by the operator transformations 
$h_{ib\uparrow}=c_{i b \downarrow}^\dagger$ and $h_{i b \downarrow}=-c_{i b \uparrow}^\dagger$, and the pair hopping interaction is neglected, ${\bm d}=(d_X, d_Y, d_Z)$ corresponds to the ``$d$ vector'' in the triplet superconductivity. 

In order to derive the effective model Hamiltonian in the low-energy sector, we use the standard canonical transformation up to the second order of ${\cal H}_t$ given by 
\begin{align}
 \left({\cal H}_{\rm eff}\right)_{\alpha\alpha'}=&
({\cal H}_U)_{\alpha \alpha'} \nonumber\\
 &+\frac{1}{2}\sum_\beta\left(
 \frac{({\cal H}_t)_{\alpha \beta} ({\cal H}_t)_{\beta \alpha'}}{E_{\alpha}-E_\beta}
 +\frac{({\cal H}_t)_{\alpha \beta} ({\cal H}_t)_{\beta \alpha'}}{E_{\alpha'}-E_\beta}
 \right) ,
\label{eq:9}
\end{align}
where $\alpha$ and $\alpha'$ belong to $\{ |H_{+1}\rangle, |H_{0}\rangle, |H_{-1}\rangle,  |L \rangle \}$, and $\beta$ belongs to the remaining higher-energy states. 
The low-energy effective Hamiltonian for the two-orbital Hubbard model is obtained as
\begin{align}
 {\cal H}_{\rm eff}=E_{0} -h_z\sum_i\tau_i^z+J_{z}\sum_{\means{ij}}\tau_i^z\tau_j^z
+J_s\sum_{\means{ij}}\bm{S}_i\cdot\bm{S}_j\nonumber\\
-J_x\sum_{\means{ij}\Gamma}\tau_{i\Gamma}^x\tau_{j\Gamma}^x
-J_y\sum_{\means{ij}\Gamma}\tau_{i\Gamma}^y\tau_{j\Gamma}^y,\label{eq:2}
\end{align}
where $\tau_i^\gamma=\sum_\Gamma \tau_{i\Gamma}^\gamma$. 
The energy parameters in this Hamiltonian are given by the parameters in the two-orbital Hubbard model, and their explicit forms are presented in Appendix~\ref{sec:exchange-parameters}.
We note that $J_s$ is positive, and signs of $J_x$ and $J_y$ reflect a sign of $t_b/t_a$.
We choose $t_b/t_a<0$ assuming a direct gap between the conduction and valence bands, which leads to $J_x>0$ and $J_y>0$.
In the case of a bipartite lattice, signs of $J_x$ and $J_y$ are changed by the transformations $\tau_i^x\to -\tau_i^x$ and $\tau_i^y\to -\tau_i^y$ on one of the sublattices.
There is a condition $|J_x|\geq |J_y|$.
The equivalent Hamiltonians were examined in Refs.~\cite{PhysRevLett.107.167403,PhysRevB.86.045137,PhysRevB.89.115134}. 

We mention the symmetry and conservation quantities in the effective Hamiltonian given by Eq.~(\ref{eq:2}). 
The total spin operators $\bm{S}_{\rm tot} (\equiv \sum_i \bm{S}_i)$ are conserved, reflecting the SO(3) symmetry in the Hamiltonian. 
When $I=0$, we have $J_x=J_y$ and $\tau^z_{\rm tot}(\equiv \sum_i \tau^z_i)$ is conserved.
This implies the U(1) symmetry on the $\tau^x-\tau^y$ plane, corresponding to the relative phase degree of freedom between the LS and HS states. 
In the case of $I \ne 0$, this U(1) symmetry is reduced to the $Z_2$ symmetry, and the Hamiltonian is invariant under the simultaneous transformation of $\tau^x_i \rightarrow -\tau_i^x$ or $\tau^y_i \rightarrow -\tau_i^y$ for all $i$. 
Therefore, the EI transition, where $\langle  \tau^x_i \rangle$ and/or $\langle \tau_i^y \rangle$ are finite, is classified as the spontaneous breaking of the $Z_2$ symmetry. 
This symmetry corresponds to the relative sign degree of freedom between the LS and HS states in the wave function: $C|L\rangle + C' |H\rangle$ and $C|L\rangle -C'|H\rangle$, where $|H \rangle$ belongs to $\{ |H_{+1}\rangle, |H_{0}\rangle, |H_{-1}\rangle \}$, and $C$ and $C'$ are complex numbers.

\section{Method}\label{sec:method}

In this section, we present formulas for the MF approximation and the generalized spin-wave approximation.
The Hamiltonian in Eq.~(\ref{eq:2}) are represented by a unified form as
\begin{align}
 {\cal H}_{\rm eff}=\sum_{i \xi}   E_{\xi}{\cal O}_{i \xi}+ \sum_{\means{ij}}\sum_\xi J_{\xi} {\cal O}_{i\xi}{\cal O}_{j\xi},
\end{align}
where a subscript $\xi$ classifies the interactions and ${\cal O}_{i \xi}$ represents the spin and PS operators at site $i$. 
We consider a bipartite lattice and introduce the MF order parameters 
$\langle {\cal O}_{\xi} \rangle_A $ and $\langle {\cal O}_{\xi} \rangle_B $ for sublattices $A$ and $B$, respectively. 
The Hamiltonian is divided into the MF term and the fluctuation term as ${\cal H}={\cal H}^{\rm MF}+\delta{\cal H}$. 
We have a MF Hamiltonian given by 
\begin{align}
 {\cal H}^{\rm MF}&=\sum_{i \xi}   E_{\xi}{\cal O}_{i \xi}
+z\sum_{i\in A}\sum_\xi J_{\xi} {\cal O}_{i\xi}\means{{\cal O}_{\xi}}_B \nonumber \\
& + z\sum_{j\in B}\sum_\xi J_{\xi} \means{{\cal O}_{\xi}}_A {\cal O}_{j\xi}
 - \frac{zN}{2}\sum_\xi J_{\xi}\means{{\cal O}_{\xi}}_A\means{{\cal O}_{\xi}}_B,
\end{align}
and a remaining term
\begin{align}
\delta {\cal H}=\sum_{\means{ij} i \in A j \in B}\sum_\xi J_{\xi} \delta{\cal O}_{i\xi}\delta{\cal O}_{j\xi} , 
\end{align}
where $z$ is a coordination number. 
We introduce $\delta{\cal O}_{i\xi}={\cal O}_{i\xi}-\means{{\cal O}_{\xi}}_C$ where $i$ belongs to the sublattice $C$. 
A set of the MFs $\{\means{{\cal O}_{\xi}}_C \}$ is obtained in ${\cal H}_{\rm MF}$ self-consistently in the numerical calculations.
The phase diagrams at zero temperature are obtained by calculating the MFs and energies in each phase.
The MF energy per site is explicitly given by 
\begin{align}
 E^{\rm MF}&=\sum_{i \xi}   \frac{E_{\xi}}{2}(\means{{\cal O}_{\xi}}_A+\means{{\cal O}_{\xi}}_B)
+ \frac{z}{2}\sum_\xi J_{\xi}\means{{\cal O}_{\xi}}_A\means{{\cal O}_{\xi}}_B , 
\end{align}
where the self-consistently obtained MFs are inserted.

The collective excitations are calculated using the generalized spin-wave method proposed in Ref.~\cite{PhysRevB.88.205110}, which is equivalent to the methods in Refs.~\cite{Onufrieva1985,Papanicolaou1988367,doi:10.1143/JPSJ.70.3076,doi:10.1143/JPSJ.72.1216,PhysRevB.60.6584,PhysRevB.88.224404}.
The fluctuation parts of the local operators are expanded by the projection operators, which are defined by the eigenstates of ${\cal H}_{\rm MF}$. 
This is given by 
\begin{align}
 \delta{\cal O}_{i\xi}=
 \sum_{mm'}X_{i}^{mm'}\bras{m;i}\delta{\cal O}_{i\xi}\kets{m';i},\label{eq:4}
\end{align}
where $\kets{m; i}$ $(m=0, 1, \cdots)$ is the $m$-th eigenstate of ${\cal H}_{\rm MF}$ at site $i$, and the projection operators are defined as $X_{i}^{mm'}=\kets{m;i}\bras{m';i}$.
Since the eigenenergy of $\kets{m; i}$ and the matrix elements $\bras{m;i}\delta{\cal O}_{i\xi}\kets{m';i}$ do not depend explicitly on $i$ but depend on the sublattice to which the site $i$ belongs, we denote them by $E^C_m$ and $\Delta{\cal O}^C_{\xi mm'}$, respectively. 

We apply the generalized Holstein-Primakoff (HP) transformation to the projection operators as 
\begin{align}
 X_{i}^{m0}=c_{im}^{\dagger}\left({\cal M}-\sum_{n=1}^{\cal N}c_{in}^{\dagger} c_{in}\right)^{1/2} ,
\end{align}
and $X_{i}^{0m}=(X_i^{m0})^\dagger$ for $m\geq 1$,
\begin{align}
 X_{i}^{mn}=c_{im}^\dagger c_{in} ,
\end{align}
for $m,n\geq 1$, and 
\begin{align}
 X_{i}^{00}={\cal M}-\sum_{n=1}^{\cal N}c_{in}^\dagger c_{in},
\end{align}
where $c_{in}^\dagger \ (c_{in})$ is the creation (annihilation) operator of the HP boson at site $i$, and $c_{in}$ takes $a_{in}$ and $b_{in}$ when the site $i$ belongs to the sublattice $A$ and $B$, respectively. 
We define ${\cal N}$ as the number of the excited states and ${\cal M} \equiv X_{i}^{00}+\sum_{n=1}^{\cal N}c_{in}^\dagger c_{in}$, where the constraint ${\cal M}=1$ is imposed at each site. 
The projection operators satisfy the commutation relations, $[X_{i}^{mn}, X_{i'}^{m'n'}]=\delta_{ii'}(X_{i}^{mn'}\delta_{m'n}-X_{i}^{m'n}\delta_{mn'})$, which are obtained by the commutation relations for the HP bosons.

By expanding the projection operators in terms of $1/{\cal M}$, the Hamiltonian ${\cal H}_{\rm eff}$ is expressed by the HP boson operators up to the quadratic order as 
\begin{align}
{\cal H}_{\rm SW}
 &=\sum_{\bm{q} n}
\Bigl(
   \Delta E_{n}^A a_{\bm{q}n}^\dagger a_{\bm{q}n} 
+ \Delta E_{n}^B b_{\bm{q}n}^\dagger b_{\bm{q}n} 
\Bigr)
\nonumber\\
&+\sum_{\bm{q}n m}  z \gamma_{\bm{q}} 
 \Bigl(
   J_{nm}^{AB}a_{\bm{q}n}^\dagger b_{-\bm{q}m}^\dagger 
+ {\widetilde J}_{nm}^{AB}a_{\bm{q}n}^\dagger b_{\bm{q}m}
+{\rm H.c.}
 \Bigr).
\label{eq:5}
\end{align}
We define $\Delta E_n^C=E_n^C-E_0^C$ and $\gamma_{\bm{q}}=z^{-1} \sum_{\bm{\rho}}e^{i\bm{q}\cdot\bm{\rho}}$, where $\bm{\rho}$ is the vector connecting NN sites.
The Fourier transforms of the HP bosons are introduced as
\begin{align}
c_{\bm{q}n}=\sqrt{\frac{2}{N}}\sum_{i\in C} c_{in}e^{-i\bm{q}\cdot\bm{r}_i} , 
\end{align}
and the summations for ${\bm q}$ in Eq.~(\ref{eq:5}) run over the first Brillouin zone for a unit cell including two sites. 
The hopping integrals of the HP bosons are given by 
\begin{align}
 J_{nm}^{AB}&=\sum_{\xi}J_\xi \Delta {\cal O}_{\xi n0}^A \Delta {\cal O}_{\xi m0}^B, 
 \\
{\widetilde J}_{nm}^{AB}&=\sum_{\xi}J_\xi \Delta {\cal O}_{\xi n0}^A \Delta {\cal O}_{\xi m0}^{B*}.
\end{align}
Using the Bogoliubov transformation~\cite{COLPA1978327}, Eq.~(\ref{eq:5}) is diagonalized as
\begin{align}
  {\cal H}_{\rm SW}=\sum_{\bm{q}\mu}\left(
 \omega_{\bm{q}\mu}^\alpha \alpha_{\bm{q}\mu}^\dagger \alpha_{\bm{q}\mu}
+\omega_{\bm{q}\mu}^\beta \beta_{\bm{q}\mu}^\dagger \beta_{\bm{q}\mu}
 \right)+{\rm const.},\label{eq:1}
\end{align}
where $\omega_{\bm{q}\mu}^\alpha$ and $\omega_{\bm{q}\mu}^\beta$ are the energies for the bosons $\alpha_{\bm{q}\mu}$ and $\beta_{\bm{q}\mu}$, respectively.
The HP boson operators and the quasi-particle operators introduced above are connected by 
${\cal B}_{\bm q}={\cal J}_{\bm{q}} {\cal A}_{\bm q}$ with  ${\cal A}_{\bm q}^\dagger=(\{a_{\bm{q}n}^\dagger\}, \{b_{\bm{q}n}^\dagger\}, \{a_{-\bm{q}n}\}, \{b_{-\bm{q}n}\})$ and ${\cal B}_{\bm q}^\dagger=(\{\alpha_{\bm{q}\mu}^\dagger\}, \{\beta_{\bm{q}\mu}^\dagger\}, \{\alpha_{-\bm{q}\mu}\}, \{\beta_{-\bm{q}\mu}\})$. 
The transformation matrix is expressed as 
\begin{align}
 {\cal J}_{\bm{q}}^{-1}=
  \begin{pmatrix}
  U_{\bm{q}} & W_{\bm{q}}\\
  V_{\bm{q}} & X_{\bm{q}}
 \end{pmatrix},
 \label{eq:matrice}
\end{align}
where $ U_{\bm{q}}$, $ V_{\bm{q}}$, $ W_{\bm{q}}$ and $ X_{\bm{q}}$ are $2{\cal N} \times 2 {\cal N}$ matrices, which diagonalize the Hamiltonian in Eq.~(\ref{eq:5})~\cite{COLPA1978327}.

Based on the generalized spin-wave method introduced above, we formulate the spin and PS excitation spectra. 
The dynamical susceptibilities at zero temperature are given as 
\begin{align}
 \chi_{\xi\xi'}(\bm{q},\omega)&=i\int_{0}^\infty dt\brasd{0}[\delta{\cal O}_{\bm{q}\xi}(t),\delta{\cal O}_{-\bm{q}\xi'}]\ketsd{0}e^{i\omega t-\epsilon t}\nonumber\\
&=-\int_{-\infty}^\infty dE\frac{{\cal S}_{\xi\xi'}({\bm q}, E)}{\omega-E+i\epsilon},\label{eq:7}
\end{align}
where $\ketsd{0}$ is the vacuum for $\alpha_{{\bm q} \mu}$ and $\beta_{{\bm q} \mu}$, ${\cal S}_{\xi\xi'}({\bm q}, E)$ is the spectral function, $\epsilon$ is an infinitesimal constant, and $\delta{\cal O}_{\bm{q}\xi}$ is defined by
\begin{align}
\delta {\cal O}_{\bm{q} \xi}=\sqrt{\frac{1}{N}}\sum_{i} \delta {\cal O}_{i \xi}e^{-i\bm{q}\cdot\bm{r}_i}.
\end{align}
The spectral function is given by
\begin{align}
 {\cal S}_{\xi\xi'}({\bm q}, E)
 &=\sum_{\bm{q}'\mu \gamma}
\brasd{0}\delta{\cal O}_{\xi\bm{q}}\ketsd{\mu;\bm{q}',\gamma}
 \brasd{\mu;\bm{q}',\gamma}\delta{\cal O}_{\xi'-\bm{q}}\ketsd{0}
\nonumber \\ &\times
\delta(E-\omega_{\bm{q}'\mu}^\gamma),
\label{eq:lehman}
\end{align}
where $\ketsd{\mu;\bm{q}, \gamma}=\gamma_{\bm{q}\mu}^\dagger\ketsd{0}$ 
and $\gamma_{{\bm q} \mu}=(\alpha_{{\bm q} \mu},  \beta_{{\bm q} \mu})$ for $\gamma=(\alpha, \beta)$. 
The spectral functions are represented using the matrices introduced in Eq.~(\ref{eq:matrice}) as 
\begin{align}
 {\cal S}_{\xi\xi'}({\bm q}, E)&=\sum_{\mu\gamma} {\cal W}_{\bm{q}\xi\mu}^{\gamma} {\cal W}_{\bm{q}\xi'\mu}^{\gamma*} \delta(E-\omega_{\bm{q} \mu}^\gamma),\label{eq:17}
\end{align}
with 
\begin{align}
{\cal W}_{\bm{q}\xi\mu}^{\gamma}=
\sum_{nC} \left(
  \Delta{\cal O}_{\xi n0}^C       [V_{\bm{q}}]_{(n,C)(\mu,\gamma)}
+\Delta{\cal O}_{\xi n0}^{C*}[U_{\bm{q}}]_{(n,C)(\mu,\gamma)}
\right) . \label{eq:18}
\end{align}
The spectral function is related to the dynamical structure factor as
\begin{align}
 {\cal S}_{\xi\xi'}(\bm{q},\omega)=\int_{-\infty}^\infty \frac{dt}{2\pi}\brasd{0}\delta{\cal O}_{\bm{q}\xi}(t)\delta{\cal O}_{-\bm{q}\xi'}\ketsd{0}e^{i\omega t}\label{eq:8}
\end{align}
for $\omega>0$.
The corresponding static susceptibilities are introduced as 
\begin{align}
\chi_{\xi\xi'} =\chi_{\xi\xi'}(\bm{q},\omega= 0)\Big|_{\bm{q}\to 0}.
\label{eq:sus}
\end{align}

We summarize a procedure to calculate the dynamical correlation functions given in Eq.~(\ref{eq:8}).
The matrix elements $\Delta{\cal O}_{\xi n0}^C$ are calculated using the MFs obtained self-consistently.
By applying the Bogoliubov transformation to the Hamiltonian in Eq.~(\ref{eq:1}), the dispersion relations $\omega_{\bm{q}\mu}^\alpha$ and $\omega_{\bm{q}\mu}^\beta$, and the transformation matrix ${\cal J}_{\bm{q}}^{-1}$ are obtained. 
Using $\Delta{\cal O}_{\xi n0}^C$, and the submatrices $U_{\bm{q}}$ and $V_{\bm{q}}$ in ${\cal J}_{\bm{q}}^{-1}$, we directly calculate ${\cal W}_{\bm{q}\xi\mu}^{\gamma}$ in Eq.~(\ref{eq:18}).
Finally, the spectral functions are obtained as 
\begin{align}
 {\cal S}_{\xi\xi'}({\bm q}, E)&=-\frac{1}{\pi}
\sum_{\mu\gamma} {\cal W}_{\bm{q}\xi\mu}^{\gamma} {\cal W}_{\bm{q}\xi'\mu}^{\gamma*} {\rm Im}\left( \frac{1}{E-\omega_{\bm{q} \mu}^\gamma+i\tilde{\epsilon}}\right ),
\label{eq:19}
\end{align}
where $\tilde{\epsilon}$ is an infinitesimal constant.

\begin{figure*}[t]
\begin{center}
\includegraphics[width=2\columnwidth,clip]{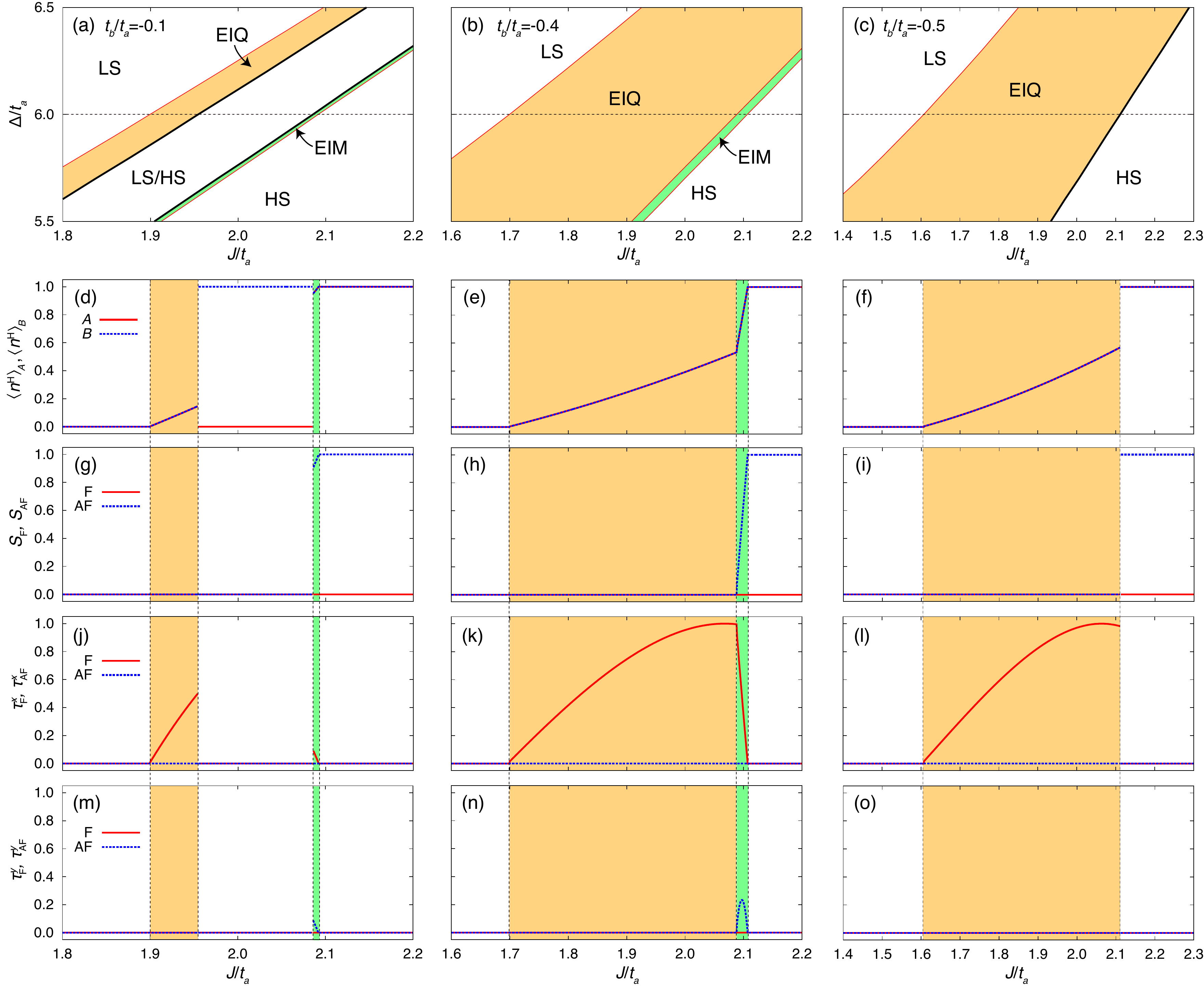}
\caption{
Phase diagrams in the planes of $J/t_a$ and $\Delta/t_a$ at (a) $t_b/t_a=-0.1$, (b) $t_b/t_a=-0.4$, and (c) $t_b/t_a=-0.5$.
The parameter values are chosen to be $I=J$, $U=6J$, and $U'=4J$.
Black thick lines and red thin lines represent the phase boundaries, where the phase transitions are of the first and second orders, respectively.
The symbols LS, HS, LS/HS, EIQ, and EIM, indicate the LS phase, AFM-HS phase, LS/HS ordered phase, EIQ phase, and EIM phase, respectively. 
The HS density ($\means{n^{\rm H}}_A$, $\means{n^{\rm H}}_B$), the spin order parameters ($S_{\rm F}$, $S_{\rm AF}$), the $x$ component of the PS order parameters ($\tau_{\rm F}^x$, $\tau_{\rm AF}^x$), and the $y$ component of the PS order parameters ($\tau_{\rm F}^y$, $\tau_{\rm AF}^y$) are plotted in (d), (g), (i), (m) for $t_b/t_a=-0.1$, (e) (h), (k), (n) for $t_b/t_a=-0.4$, and (f), (i), (l), (o)  for $t_b/t_a=-0.5$. 
 In (d)--(o), we choose $\Delta/t_a=6.0$ indicated by the horizontal broken lines in (a), (b) and (c). 
}
\label{mf}
\end{center}
\end{figure*}

\section{Ground state phase diagram}\label{sec:ground-state-phase}

\subsection{MF phase diagram}

In this subsection, we present the numerical results of the ground-state phase diagrams obtained by the MF approximation. 
All numerical calculations are performed in the two-dimensional square lattice with the coordination number $z=4$. 
First, we show the phase diagram on the plane of $\Delta$ and $J$ for several values of $t_b/t_a$ in Figs.~\ref{mf}(a)--(c). 
The phase boundaries are determined by calculating the spin and PS order parameters. 
The spin states are identified by $\means{n^{\rm H}}_C$, which takes 0 and 1 in the case where the spin states at all sites are LS and HS, respectively. 
The spin structure is examined by the square of the magnetic order parameters defined as 
\begin{align}
 S_{\rm F(AF)}=\frac{1}{4}\sum_{\gamma=x,y,z}\Bigl(\means{S^\gamma}_A\pm\means{S^\gamma}_B\Bigr)^2,
 \label{eq:mag}
\end{align}
where $+$ and $-$ in the right hand side imply the ferromagntic and antiferromagnetic orders, respectively. 
The EI order is identified by the $x$ and $y$ components of the PS operators defined by
\begin{align}
 \tau_{\rm F(AF)}^\gamma=\frac{1}{4}
\sum_{\Gamma=X,Y,Z}
\Bigl(\means{\tau_{\Gamma}^\gamma}_A\pm\means{\tau_{\Gamma}^\gamma}_B\Bigr)^2,
\end{align}
where $\gamma$ takes $x$ and $y$.

Let us start from Fig.~\ref{mf}(a) for $t_b/t_a=-0.1$. 
The order parameters as functions of $J/t_a$ are presented in Figs.~\ref{mf}(d), \ref{mf}(g), \ref{mf}(j) and \ref{mf}(m) at $\Delta/t_a=6$, which is plotted by a dotted horizontal line in Fig.~\ref{mf}(a). 
The five phases appear at this parameter. 
In the case of $\Delta\gg J$, we have $\means{n^{\rm H}}_A=\means{n^{\rm H}}_B=0$, implying the LS phase. 
On the other side, in the case of $\Delta \ll J$, $\means{n^{\rm H}}_A=\means{n^{\rm H}}_B=1$, implying the HS phase.
The magnetic order parameters shown in Fig.~\ref{mf}(g) indicate that the $S=1$ spins are aligned antiferromagnetically.
This phase is termed AFM-HS phase from now on, and is mainly attributed to the exchange interaction in the fourth term in Eq.~(\ref{eq:2}).
The wave functions on the two sublattices in this phase are given as
 \begin{align}
 \kets{\psi_{\rm HS}}_A&=\kets{H_{+1}},\\ 
\kets{\psi_{\rm HS}}_B&=\kets{H_{-1}}.
 \end{align}
Between the LS and AFM-HS phases, we find a phase which is characterized by $\means{n^{\rm H}}_A=0$ and $\means{n^{\rm H}}_B=1$. 
 This is the spin-state ordered phase denoted as LS/HS, where the HS and LS sites are aligned alternately in the square lattice.  
 These three phases, the LS, AFM-HS and LS/HS ordered phases, in the two-orbital Hubbard model have been shown in Refs.~\cite{PhysRevLett.107.167403,PhysRevB.86.045137,PhysRevB.80.024423}.
Note that the spin alignments in the LS/HS ordered phase are not fixed in the present MF approximation, and the long-range exchange interactions are necessary to determine the spin structure. 

We find two phases located between the LS and LS/HS ordered phase, and between the AFM-HS and LS/HS ordered phase as shown in Fig.~\ref{mf}(a).
In both the two phases, $\means{n^{\rm H}}_A$ and $\means{n^{\rm H}}_B$ take values between 0 and 1, implying that the LS and HS states are mixed quantum mechanically. 
Since there are no direct hopping integrals between the $a$ and $b$ orbitals, this mixing occurs spontaneously due to the interactions. 
This is directly confirmed by the $x$ and $y$ components of the PS order parameters shown in Figs.~\ref{mf}(j) and~\ref{mf}(m). 
Both the two phases are identified as the EI phases. 

We focus on the EI phase appearing between the LS and LS/HS ordered phases.
We term this phase EIQ.
It is shown that $\tau_{\rm F}^x$ is only finite among the several PS order parameters, implying that this is a uniform EI phase with the real wave function.  
Any magnetic order parameters introduced in Eq.~(\ref{eq:mag}) do not emerge as shown in Fig.~\ref{mf}(g), i.e., no conventional magnetic long-range order. 
Without loss of generality, we set the order parameters in this phase as $\means{\tau_Z^x}_C \ne 0$ and $\means{\tau_X^x}_C=\means{\tau_Y^x}_C=0$, implying the wave function given by
\begin{align}
 \kets{\psi_{\rm EIQ}}=C_1 \kets{L}+C_2 \kets{H_Z}
\label{eq:12} , 
\end{align}
with real numbers $C_1$ and $C_2$. 
This type of the uniform EI order originates from the ferro-type exchange interactions $J_x>J_y>0$ in Eq.~(\ref{eq:2}). 
In this wave function, we have that $\means{S^x}=\means{S^y}=\means{S^z}=0$ and $\means{Q^{3z^2-r^2}} \ne 0$ where $Q^{3z^2-r^2} \equiv 3(S^z)^2-S(S+1)$ is one of the spin-quadrupole operators. 
This phase is identified as a spin-nematic ordered phase~\cite{PhysRevLett.97.087205,doi:10.1143/JPSJ.75.083701}, and is termed the EIQ phase from now on.

Next, we focus on the EI phase between the AFM-HS and the LS/HS ordered phases.
The finite PS moments shown in Fig.~\ref{mf}(j) and \ref{mf}(m) imply the quantum mixing of the LS and HS states.
The AFM order is realized as shown in Fig.~\ref{mf}(g), in contrast to the EIQ phase. 
We term this phase EIM.
% Several characteristics in the EIM phase are understood from the explicit forms of the wave functions.
% Without loss of generality, the staggered magnetic moment is assumed to be parallel to $S^z$,  
% and the wave functions in the two sublattices are given as 
From the numerical results, we give explicit forms of the wave functions as 
\begin{align}
 \kets{\psi_{\rm EIM}}_A&
=D_1\kets{L}+\tilde{D}_2 e^{-i\theta_A}\kets{H_{+1}}+\tilde{D}_3 e^{i\theta_A}\kets{H_{-1}},\label{eq:16}\\
 \kets{\psi_{\rm EIM}}_B&
=D_1\kets{L}+\tilde{D}_2 e^{i\theta_B}\kets{H_{-1}}+\tilde{D}_3 e^{-i\theta_B}\kets{H_{+1}},\label{eq:15}
\end{align}
where $D_1$, $\tilde{D}_2$, and $\tilde{D}_3$ are real number. Without loss of generality, the staggered magnetic moment is assumed to parallel to $S^z$, and $\theta_A$ and $\theta_B$ are phases corresponding to the spin rotation around $S^z$.
% where $D_1$, $\tilde{D}_2$, and $\tilde{D}_3$ are real numbers, and $\theta_A$ and $\theta_B$ are the phases reflecting rotations along $S^z$.
These are also expressed as 
\begin{align}
 \kets{\psi_{\rm EIM}}_A&
=D_1\kets{L}+D_2\kets{H_X}+D_3\kets{H_Y},\label{eq:13}\\
 \kets{\psi_{\rm EIM}}_B&
=D_1\kets{L}+D_2^*\kets{H_X}+D_3^*\kets{H_Y},\label{eq:14}
\end{align}
where $D_2=(-\tilde{D}_2 e^{-i\theta}+\tilde{D}_3 e^{i\theta})/\sqrt{2}$ and $D_3=-i(\tilde{D}_2 e^{-i\theta}+\tilde{D}_3 e^{i\theta})/\sqrt{2}$, and $\theta_A=\theta$ and $\theta_B=\theta +\pi$ are imposed.
Then, we have $\means{\tau_Y^x}_A/\means{\tau_X^x}_A=-\means{\tau_X^y}_A/\means{\tau_Y^y}_A=\tan\theta$, $\means{\tau_\Gamma^x}_A=\means{\tau_\Gamma^x}_B$, and $\means{\tau_{\Gamma}^y}_A=-\means{\tau_\Gamma^y}_B$ for $\Gamma=X,Y$, and $\means{\tau_Z^\gamma}_A=\means{\tau_Z^\gamma}_B=0$ for $\gamma=x,y$.
These relations imply the canted PS order associated with the AFM order, which are consistent with the results in Figs.~\ref{mf}(j) and \ref{mf}(m).

We show that a difference between the widths of the two bands controls stabilities of the EI phases. 
In Fig.~\ref{mf}(b) and \ref{mf}(c), the phase diagrams for $t_b/t_a=-0.4$ and $-0.5$ are presented, respectively. 
The detailed order parameters are presented in  Fig.~\ref{mf}. 
With increasing $|t_b/t_a|$ [Fig.~\ref{mf}(b)], the LS/HS ordered phase disappears and the two EI phases touch directly with each other, and the EIQ phase rather than the EIM phase expands.
The order parameters change continuously at the boundary between the two EI phases, implying the second order phase transition. 
Detailed discussion for characters of the phase transitions will be presented in Sec.~\ref{sec:symm-analys-phase}.
% The region of the EIQ phase rather than the EIM phase expands with increasing $|t_b/t_a|$. 
With further increasing $|t_b/t_a|$ [Fig.~\ref{mf}(c)], the EIM phase disappears and the EIQ phase fills a parameter region between the AFM-HS and LS phases. 
The calculated results for the order parameters indicate that the phase transition between the EIQ and LS (AFM-HS) phases are of the second (first) order.

\begin{figure}[t]
\begin{center}
\includegraphics[width=\columnwidth,clip]{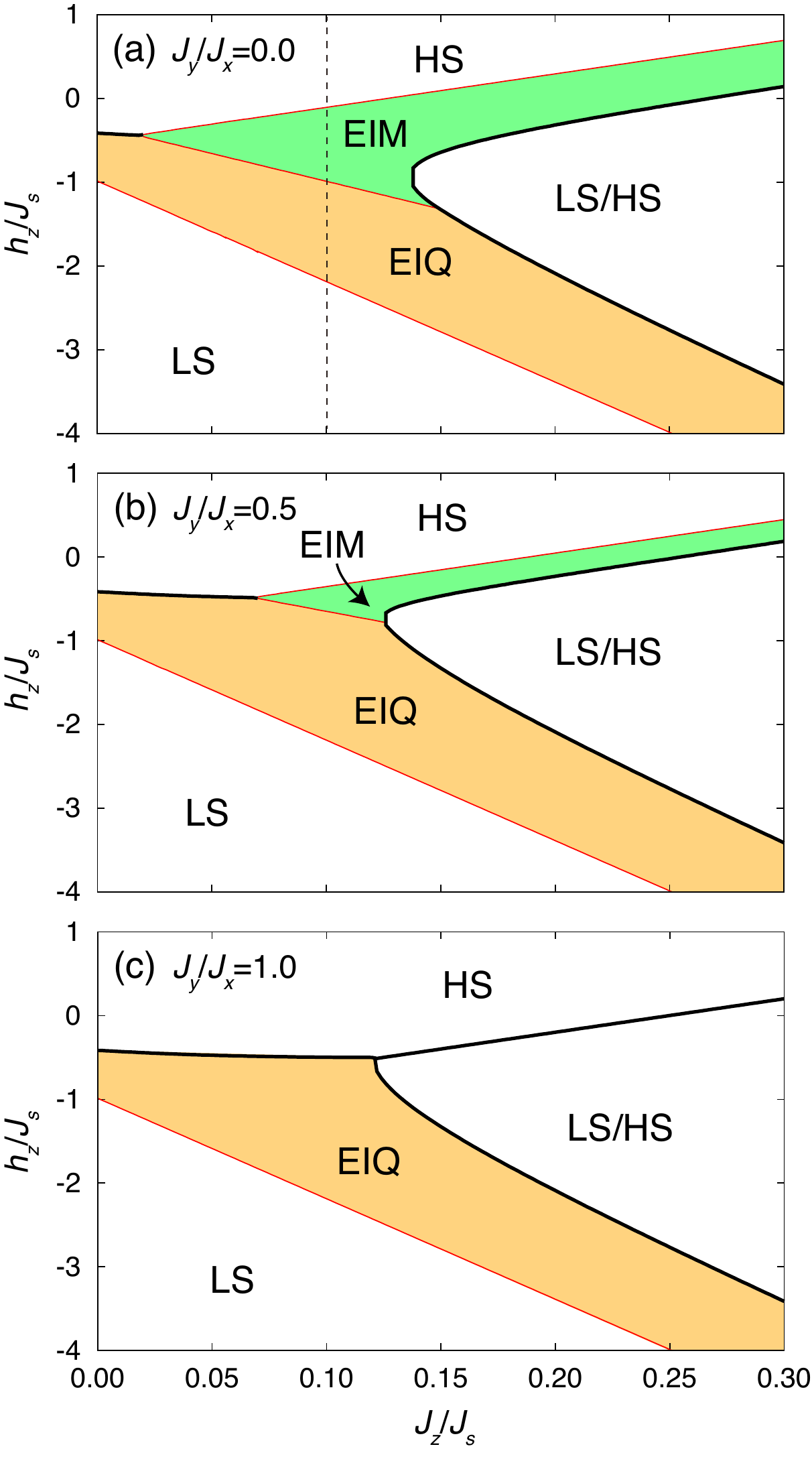}
\caption{
Phase diagrams in the planes of $J_z/J_s$ and $h_z/J_s$ for (a) $J_y/J_x=0$, (b) $J_y/J_x=0.5$, and (c) $J_y/J_x=1$.
The other parameter value is chosen to be $J_x/J_s=0.5$.
The black thick lines and red thin lines represent the phase boundaries, where the phase transitions are of the first and second orders, respectively.
The symbols LS, HS, LS/HS, EIQ, and EIM indicate the LS phase, AFM-HS phase, LS/HS ordered phase, EIQ phase, and EIM phase, respectively. 
The broken line in (a) represents the parameter region, where the quantities presented in Fig.~\ref{mf_val_dep} are calculated. 
}
\label{mf_val}
\end{center}
\end{figure}

\begin{figure}[t]
\begin{center}
\includegraphics[width=\columnwidth,clip]{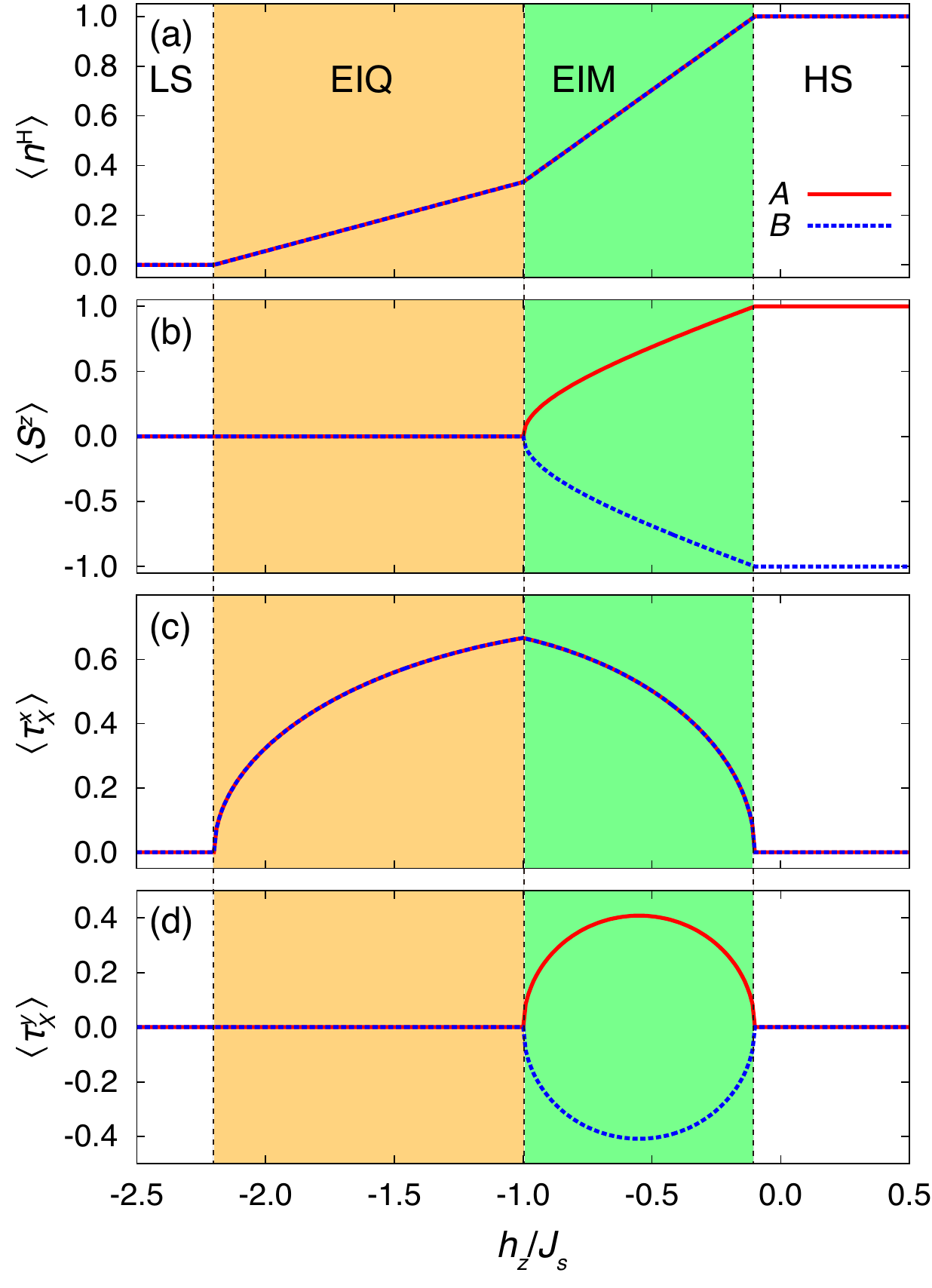}
 \caption{
 (a) High-spin density ($\means{n^{\rm H}}_A$, $\means{n^{\rm H}}_B$),  
(b) spin order parameters ($\means{S^z}_A$, $\means{S^z}_B$),
(c) the $x$ component of the PS order parameters ($\means{\tau^{x}}_A$, $\means{\tau^{x}}_B$), and
(d) the $y$ component of the PS order parameters ($\means{\tau^{y}}_A$, $\means{\tau^{y}}_B$). 
The other parameter values are chosen to be $J_z/J_s=0.1$, $J_x/J_s=0.5$ and $J_y=0$.  
Corresponding parameter region is denoted by the broken line in Fig.~\ref{mf_val}(a).
}
\label{mf_val_dep}
\end{center}
\end{figure}

In order to clarify roles of the interaction terms in the effective Hamiltonian in Eq.~(\ref{eq:2}), we show the phase diagrams on a plane of the parameters in the effective Hamiltonian. 
Among the five energy parameters, $J_s$ is chosen to be a unit of energy and $J_x/J_s$ is fixed to be 0.5.
We show the phase diagrams on a plane of $h_z/J_s$ and $J_z/J_s$ for several values of $J_y/J_x$ in Fig.~\ref{mf_val}.
Note that $h_z$ reflects the energy difference between the $a$ and $b$ orbitals and $J_z$ provides the attractive interaction between the NN LS and HS states.  
The spin-quantization axis is chosen to be $S^z$, and the conditions $\means{\tau_X^x}=\means{\tau_Y^x}$, $\means{\tau_X^y}=-\means{\tau_Y^y}$, and  $\means{\tau_Z^x}=\means{\tau_Z^y}=0$ are imposed.
It is shown that positive and negative $h_z$ stabilize the AFM-HS and LS phases, respectively, and the LS/HS ordered phase is realized with increasing $J_z$. 
Between these three phases, the two EI phases appear. 
With increasing $J_y/J_x$ from zero, the EIM phase is shrunken, and vanishes at $J_y/J_x=1$. 
As mentioned in Sec.~\ref{sec:model}, $|J_y/J_x| \leq 1$ is satisfied, and the U(1) symmetry exists in the effective Hamiltonian at $J_y/J_x=1$, which corresponds to $I=0$. 
This implies that the loss of the U(1) symmetry in the case of $I\ne 0$ is necessary to realize the EIM phase.
Detailed $h_z/J_s$ dependences of the order parameters are presented in Fig.~\ref{mf_val_dep} along the broken line in Fig.~\ref{mf_val}(a).
All order parameters are changed continuously, implying that all phase transitions in this case are of the second order.  
The orders of the phase transitions are represented by thin and bold lines in Fig.~\ref{mf_val}.

\subsection{Symmetry analysis}\label{sec:symm-analys-phase}
\begin{table}[t]
\centering
\begin{tabular}{@{\hspace{0.3em}} c @{\hspace{0.3em}}| @{\hspace{0.75em}} c @{\hspace{1.5em}} c @{\hspace{1.5em}} c @{\hspace{1.5em}} c @{\hspace{1.5em}} c @{\hspace{1.5em}} c @{\hspace{1.5em}} c @{\hspace{0.75em}}}
  &  $\Theta$ & ${\cal T}$ & ${\cal G}$ & $\Theta{\cal T}$ & $\Theta{\cal G}$ & ${\cal TG}$ &  $\Theta{\cal TG}$ \\
  \hline
  \hline
  LS    & Yes & Yes & Yes & Yes & Yes & Yes & Yes \\
  \hline
  EIQ   & No  & Yes & No  & No  & Yes & No  & Yes \\
  \hline
  EIM   & No  & No  & No  & No  & No  & No  & Yes \\
  \hline
  HS    & No  & No  & Yes & Yes & No  & No  & Yes \\
  \hline
  LS/HS & Yes & No  & Yes & No  & Yes & No  & No  \\
  \hline
 \end{tabular}
 \caption{
Symmetries of the phases which appear in the ground-state phase diagram.
The symbols LS, HS, LS/HS, EIQ, and EIM indicate the LS phase, the AFM-HS phase, the LS/HS ordered phase, the EIQ phase, and the EIM phase, respectively. 
The symbols ${\cal T}$, ${\cal G}$, and $\Theta$ are the generators for the translational symmetry, the $Z_2$ symmetry for PS, and time-reversal symmetry, respectively.
``Yes'' and ``No'' imply that the phase is invariant and is not invariant under the corresponding operation.
Any magnetic orders are not assumed in the LS/HS ordered phase.
}
\label{symm_tab}
\end{table}

We analyze the order of the phase transition from the symmetry viewpoint. 
In Table~\ref{symm_tab}, the symmetries in the five phases in the ground state are summarized.
Each phase is characterized by the following four kinds of the symmetries; the SO(3) symmetry for the spin rotation, the time-reversal symmetry, the translational symmetry, and the $Z_2$ symmetry in terms of the PS inversion in the $\tau^x-\tau^y$ plane given as $\tau^x \rightarrow -\tau^x$ and $\tau^y \rightarrow -\tau^y$.
The generators of the time-reversal symmetry, the translational symmetry, and the PS-reversal symmetry are denoted by $\Theta$, ${\cal T}$, and ${\cal G}\equiv \exp[i\pi\sum_i(n_i^{\rm H}-n_i^{\rm L})/2]$, respectively.
We have the relations $\Theta C\kets{L}=C^*\kets{L}$, ${\cal G} C\kets{L}=C\kets{L}$,  $\Theta C\kets{H_\Gamma}=-C^*\kets{H_\Gamma}$, and ${\cal G} C\kets{H_\Gamma}=-C\kets{H_\Gamma}$, where $C$ is a complex number. 
It is clear that the LS phase retains the all symmetries, and the AFM-HS state is invariant under the operations ${\cal G}$ and $\Theta{\cal T}$. 
The LS/HS ordered phase breaks the translational symmetry.
For the EI phases, we have $\Theta {\cal G}\kets{\psi_{\rm EIQ}}=\kets{\psi_{\rm EIQ}}$ and $\Theta {\cal G}\kets{\psi_{\rm EIM}}_A=\kets{\psi_{\rm EIM}}_B$, where $\kets{\psi_{\rm EIQ}}$, $\kets{\psi_{\rm EIM}}_A$ and $\kets{\psi_{\rm EIM}}_B$ are defined in Eqs.~(\ref{eq:12}), (\ref{eq:13}) and~(\ref{eq:14}), respectively.
Hence, the EIQ phase is invariant under the operations of $\Theta{\cal G}$ and ${\cal T}$, and the EIM is invariant under $\Theta{\cal TG}$.

Through the symmetry analyses listed in Table~\ref{symm_tab}, the phase boundaries are classified by the Landau's criteria for the second order phase transition. 
Since there are the inclusion relations between the symmetries in the LS and EIQ phases, those between the EIQ and EIM phases, and between the EIM and AFM-HS phases, the phase transitions are possible to be of the second order. 
On the other hand, since there is not the inclusion relations between the symmetries in the EIQ and AFM-HS phases, and between the LS/HS ordered phase and the other phases except for the LS phase, the phase transitions between these phases should be of the first order. 
The symmetry considerations shown above are consistent with the numerical results presented in Fig.~\ref{mf}. 
There are possibilities at finite temperatures that the LS phase directly borders the LS/HS ordered phase as well as the AFM-HS phase across the boundaries with the second-order phase transition.

\section{Collective excitations}\label{sec:coll-excit}
\subsection{Dispersion relation and excitation spectra}
\label{sec:dis}

\begin{figure*}[t]
\begin{center}
\includegraphics[width=2\columnwidth,clip]{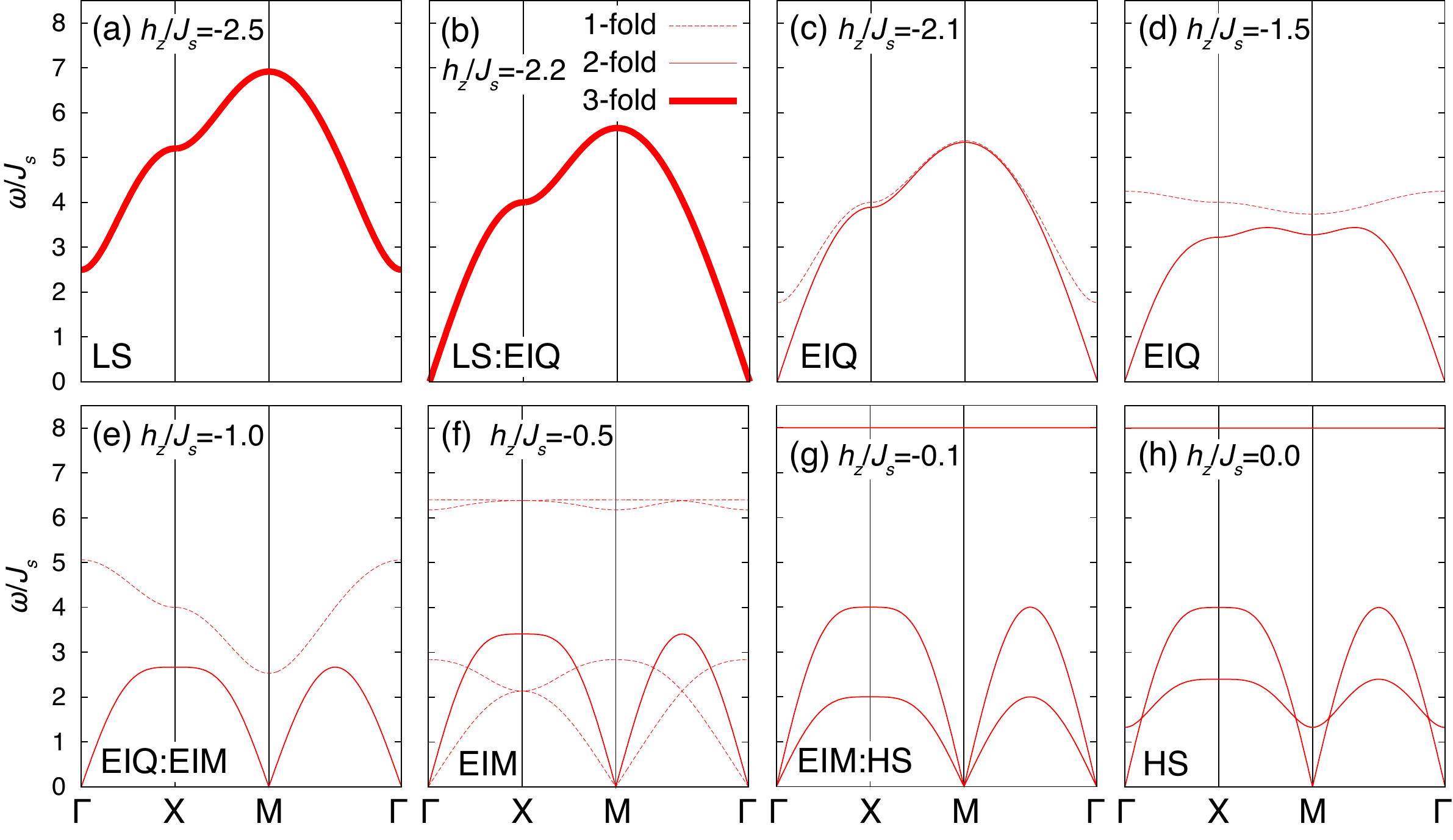}
 \caption{
 Dispersion relations of the collective modes at (a) $h_z/J_s=-2.5$, (b) $h_z/J_s=-2.2$, (c) $h_z/J_s=-2.1$, (d) $h_z/J_s=-1.5$, (e) $h_z/J_s=-1.0$, (f) $h_z/J_s=-0.5$, (g) $h_z/J_s=-0.1$, and (h) $h_z/J_s=0$. 
The other parameter values are chosen to be $J_z/J_s=0.1$, $J_x/J_s=0.5$ and $J_y=0$. 
The number of the degeneracy is indicated by types of the lines.
The symbols LS, HS, LS/HS, EIQ, and EIM, respectively, indicate the LS phase, the AFM-HS phase, the LS/HS ordered phase, the EIQ phase, and the EIM phase, respectively.
The horizontal axes are unified in (a)-(h), although the volume of the magnetic unit cell in (f)-(h) are twice of that in (a)-(e). 
Dispersion relations are plotted in the first Brillouin zone for a unit cell, where the bond lengths connecting the NN sites are taken as a lattice constant. 
The so-called reduced and periodic zone pictures are adopted in (a)-(e), and (f)-(h), respectively. 
The $\Gamma$ and M points are equivalent with each other in (f)-(h). 
}
\label{band}
\end{center}
\end{figure*}

In this section, we present the numerical results of the collective excitations.
First, we present the dispersion relations of the collective excitations obtained by the spin-wave approximation in Fig.~\ref{band}, where $h_z$ is changed along a dotted line in Fig.~\ref{mf_val}(a). 
Other parameter values are the same with those in Fig.~\ref{mf_val_dep}. 
The six excitation modes are attributed to the three excitations per site in the two sublattices, i.e., the spin excitation, orbital excitation, and simultaneous spin-orbital excitation.

The results in the LS phase ($h_z/J_s=-2.5$) are presented in Fig.~\ref{band}(a). 
In an isolated LS state,  the excited states are the triply-degenerate HS states. 
The dispersions are caused by the interactions $J_x\sum_{\means{ij}\Gamma}\tau_{i\Gamma}^x\tau_{j\Gamma}^x$ and $J_y\sum_{\means{ij}\Gamma}\tau_{i\Gamma}^y\tau_{j\Gamma}^y$ in the effective Hamiltonian. 
The center of the excitation bands is located around $2|h_z|$ which is the energy gap between the LS and HS states in the case without the intersite interactions. 
At the phase boundary between the LS and EIQ phase ($h_z/J_s=-2.2$), one of the two bands touches the zero energy at the Brillouin-zone center as shown in Fig.~\ref{band}(b).
On the other side, in the AFM-HS phase shown in Fig.~\ref{band}(h), the doubly-degenerate Goldstone modes are identified as the spin waves in the AFM ordered state.

\begin{figure}[t]
\begin{center}
\includegraphics[width=\columnwidth,clip]{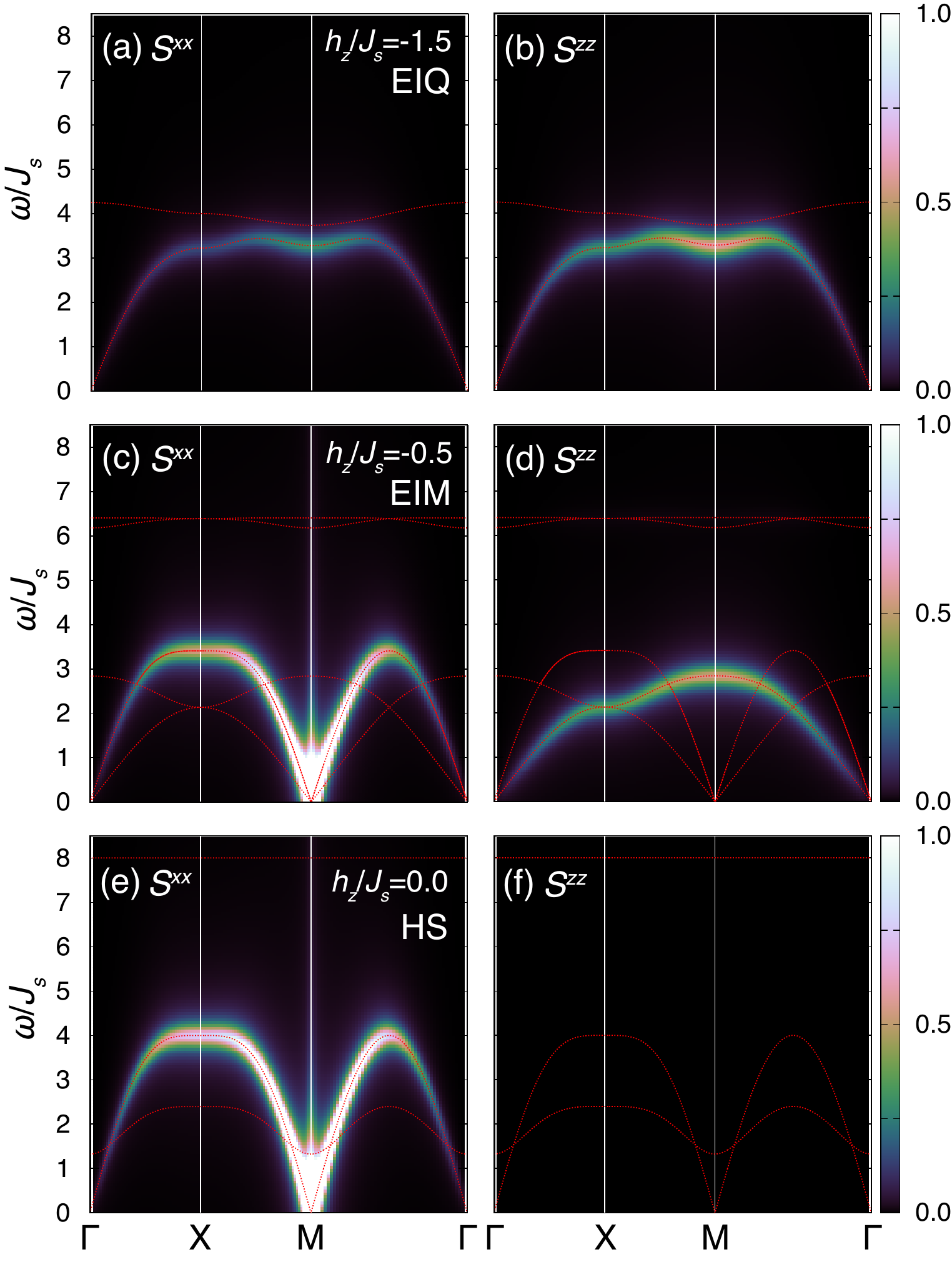}
 \caption{
Contour maps of the dynamical spin correlation functions. 
The $x$ components $S^{xx}$ and the $z$ components $S^{zz}$ are shown in (a), (c), (e), and (b), (d), (f), respectively. 
The parameter value of $h_z/J_s$ is taken to be (a) $-1.5$ (the EIQ phase), (b) $-0.5$ (the EIM phase) and (c) 0 (the AFM-HS phase).
The horizontal axes are taken in the same manner as Fig.~\ref{band}.
A value of the constant in the denominator in Eq.~(\ref{eq:19}) is chosen to be $\tilde{\epsilon}/J_s=0.2$.
The other parameter values are the same with those in Fig.~\ref{mf_val_dep}.
The dotted red lines represent the dispersion relations of the collective modes [see Figs.~\ref{band}(d), \ref{band}(f), and ~\ref{band}(h)].
Intensities of $S^{zz}$ vanish in (f).
}
\label{spec}
\end{center}
\end{figure}

\begin{figure}[t]
\begin{center}
\includegraphics[width=\columnwidth,clip]{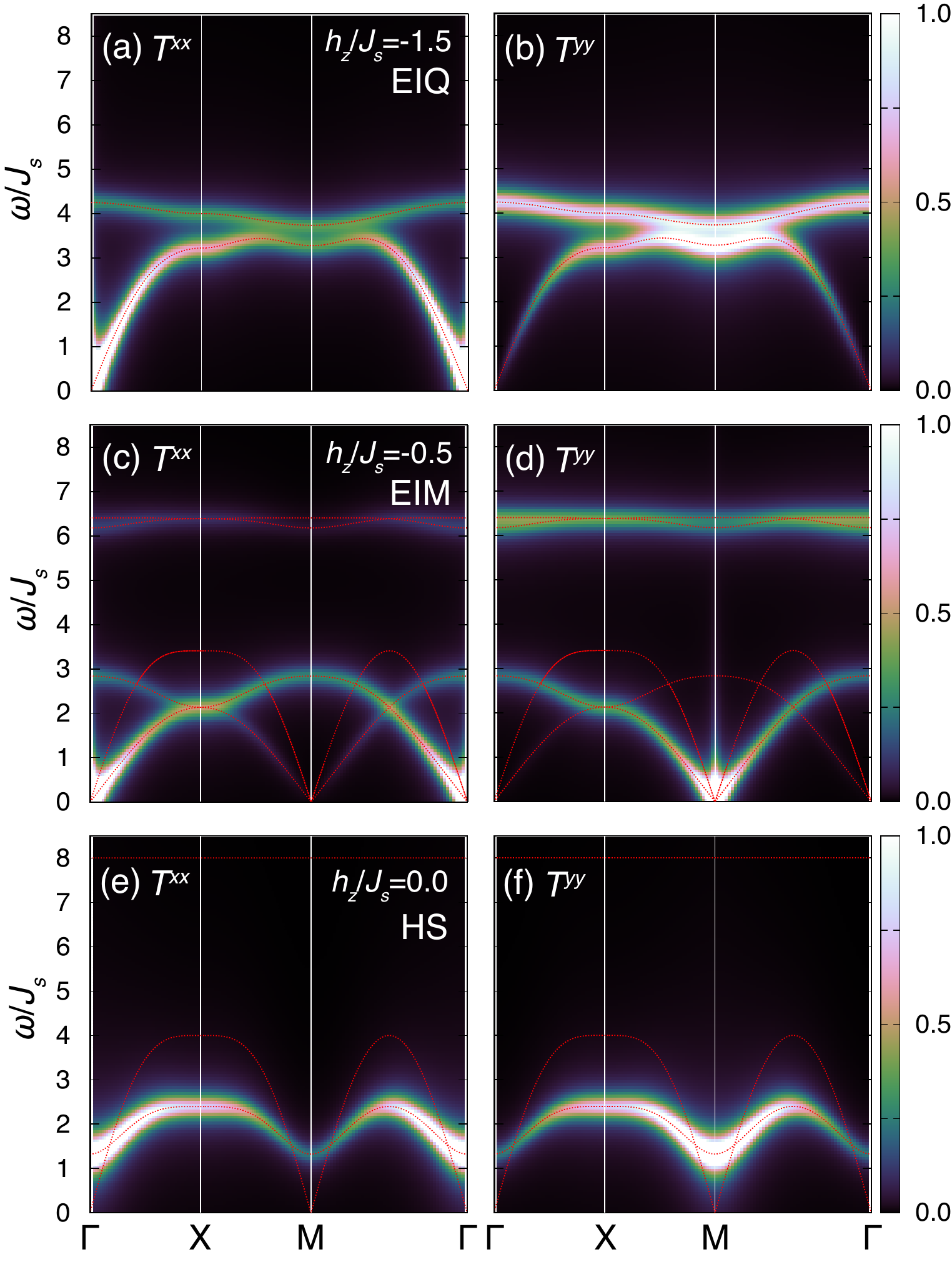}
 \caption{
 Contour maps of the dynamical PS correlation functions. 
The $x$ components $T^{xx}$ and the $y$ components $T^{yy}$ are shown in (a), (c), (e), and (b), (d), (f), respectively. 
The parameter value of $h_z/J_s$ is taken to be (a) $-1.5$ (the EIQ phase), (b) $-0.5$ (the EIM phase) and (c) 0 (the AFM-HS phase). 
The horizontal axes are taken in the same manner as Fig.~\ref{band}.
A value of the constant in the denominator in Eq.~(\ref{eq:19}) is chosen to be $\tilde{\epsilon}/J_s=0.2$.
The other parameter values are the same with those in Fig.~\ref{mf_val_dep}.
The dotted red lines represent the dispersion relations of the collective modes [see Figs.~\ref{band}(d), \ref{band}(f), and ~\ref{band}(h)].
}
\label{spec_orb}
\end{center}
\end{figure}

The results in the EIQ phase are shown in Figs.~\ref{band}(c) and~\ref{band}(d).
The triply-degenerate lower-energy bands shown in the LS phase split into the doubly-degenerate Goldstone modes and the non-degenerate gapful mode. 
The former is the spin-wave excitations in the spin-nematic phase~\cite{PhysRevLett.97.087205,doi:10.1143/JPSJ.75.083701}.
The latter corresponds to the excitation which changes a relative weight of the LS and HS states in the wave function, as discussed later in more detail [see Figs.~\ref{spec}(a) and~\ref{spec}(b) and Fig.~\ref{spec_orb}(a) and~\ref{spec_orb}(b)]. 
At the boundary between the EIQ and EIM phases ($h_z/J_s=-1.0$) shown in Fig.~\ref{band}(e), the Goldstone modes are fourfold degenerate.
In the EIM phase, the two kinds of the Goldstone modes appear; 
the doubly-degenerate modes with higher velocity and the non-degenerate mode with lower velocity. 
As shown later in more detail [see Figs.~\ref{spec}(c) and~\ref{spec}(d) and Figs.~\ref{spec_orb}(c) and~\ref{spec_orb}(d)], 
the former and latter correspond to the transverse- and longitudinal-magnetic excitations, respectively.  

Characteristics in the collective modes are examined by calculating the dynamical correlation functions for the spin and PS operators introduced in Eq.~(\ref{eq:8}).
The numerical results are presented in Figs.~\ref{spec} and \ref{spec_orb}.
The results in the EIQ, EIM, and AFM-HS phases are calculated by changing $h_z/J_s$, and other parameter values are  the same with those in Fig.~\ref{mf_val_dep}. 
For simplicity, we introduce a notation $S^{\gamma\gamma}(\bm{q},\omega)$ ($\gamma=x,y,z$) as the dynamical spin correlation function, i.e., ${\cal S}_{\xi\xi}(\bm{q},\omega)$ introduced in Eq.~(\ref{eq:8}) with Eq.~(\ref{eq:lehman}), where $\delta {\cal O}_{\bm{q} \xi}$ is taken to be $S^\gamma_{\bm{q}}-\means{S_{\bm{q}}^\gamma}$.
In the same way, we introduce a notation $T_\Gamma^{\gamma\gamma}(\bm{q},\omega)$ as the dynamical PS correlation functions for the operator $\tau^\gamma_{\bm{q}\Gamma}-\means{\tau_{\bm{q}\Gamma}^\gamma}$.
We have the following relations: $S^{xx}(\bm{q},\omega)=S^{yy}(\bm{q},\omega)$, $T^{xx}(\bm{q},\omega)\equiv T_X^{xx}(\bm{q},\omega)=T_Y^{xx}(\bm{q},\omega)$ and $T^{yy}(\bm{q},\omega) \equiv T_X^{yy}(\bm{q},\omega)=T_Y^{yy}(\bm{q},\omega)$, when $S^z$ is taken to be the quantization axis.
The relations $\means{\tau_X^x}=\means{\tau_Y^x}$, $\means{\tau_X^y}=-\means{\tau_Y^y}$, and $\means{\tau_Z^x}=\means{\tau_Z^y}=0$ are imposed.

First, we focus on the dynamical spin correlation functions, $S^{xx}(\bm{q},\omega)$ and $S^{zz}(\bm{q},\omega)$, which correspond to the transverse and longitudinal excitations, respectively, shown in Fig.~\ref{spec}. 
In the EIQ phase [Figs.~\ref{spec}(a) and~\ref{spec}(b)], the doubly-degenerate % Goldstone
lower-energy collective modes contribute to both the transverse and longitudinal excitation spectra.
This fact permits us to identify experimentally the EIQ phase where conventional magnetic orders do not emerge.
The intensity vanishes around the $\Gamma$ point, because of no long-range orders in the spin sectors~\cite{PhysRevLett.97.087205}.

In the EIM phase [Figs~\ref{spec}(c) and~\ref{spec}(d)], as mentioned previously, 
there appear the two kinds of the Goldstone modes, i.e., the doubly-degenerate modes with a higher velocity and the non-degenerate mode with a lower velocity. 
The formers are identified as the transverse excitations, which originate from the spin-wave excitations in the AFM order. 
The latter is the longitudinal component attributed to change in magnitude of the spin moment. 
This implies a change in a relative weight of the LS and HS states in the wave function. 
In contrast to the results in the two EI phases, the spin excitation spectra in the HS phase shown in Figs.~\ref{spec}(e) and~\ref{spec}(f) have only the transverse components originating from the spin-wave excitations in the AFM order. 
As shown in Figs.~\ref{spec}(c) and~\ref{spec}(e), the intensities in $S^{xx}$ are remarkable around the M point in the EIM and HS phases, owing to the AFM order.

The dynamical PS correlation functions are shown in Fig.~\ref{spec_orb}. 
In the EIQ phase [Figs.~\ref{spec_orb}(a) and~\ref{spec_orb}(b)], both the optical modes located around $\omega/J_s=4$ and the Goldstone modes show finite spectral intensities. 
Since the formers do not contribute to the dynamical spin correlation functions as shown in Figs.~\ref{spec}(a) and~\ref{spec}(b), these are the pure orbital excitations. 
The Goldstone mode is observed in $T^{xx}$ because of the $\tau^x_{\rm F}$ order.
In the EIM phase shown in Figs.~\ref{spec_orb}(c) and~\ref{spec_orb}(d), the optical modes located around $\omega/J_s=6$, and the Goldstone modes provide finite weights in the spectra. 
Intensities in $T^{xx}$ ($T^{yy}$) are remarkable around the $\Gamma$ (M) point 
due to the $\tau_{\rm F}^x$ ($\tau_{\rm AF}^y$) order in this phase.
The results in the AFM-HS phase are shown in Figs.~\ref{spec_orb}(e) and~\ref{spec_orb}(f). 
The spectral intensities for the optical modes located around $\omega/J_s=8$ are zero, since these excitations are accompanied by changing $S^z$ by $\pm 1$, i.e., the spin-nematic excitations.
Large spectral intensity are seen at the optical modes $\omega/J_s=1-2$, identified as the pure orbital excitations from the HS to LS states.

\begin{figure}[t]
\begin{center}
\includegraphics[width=\columnwidth,clip]{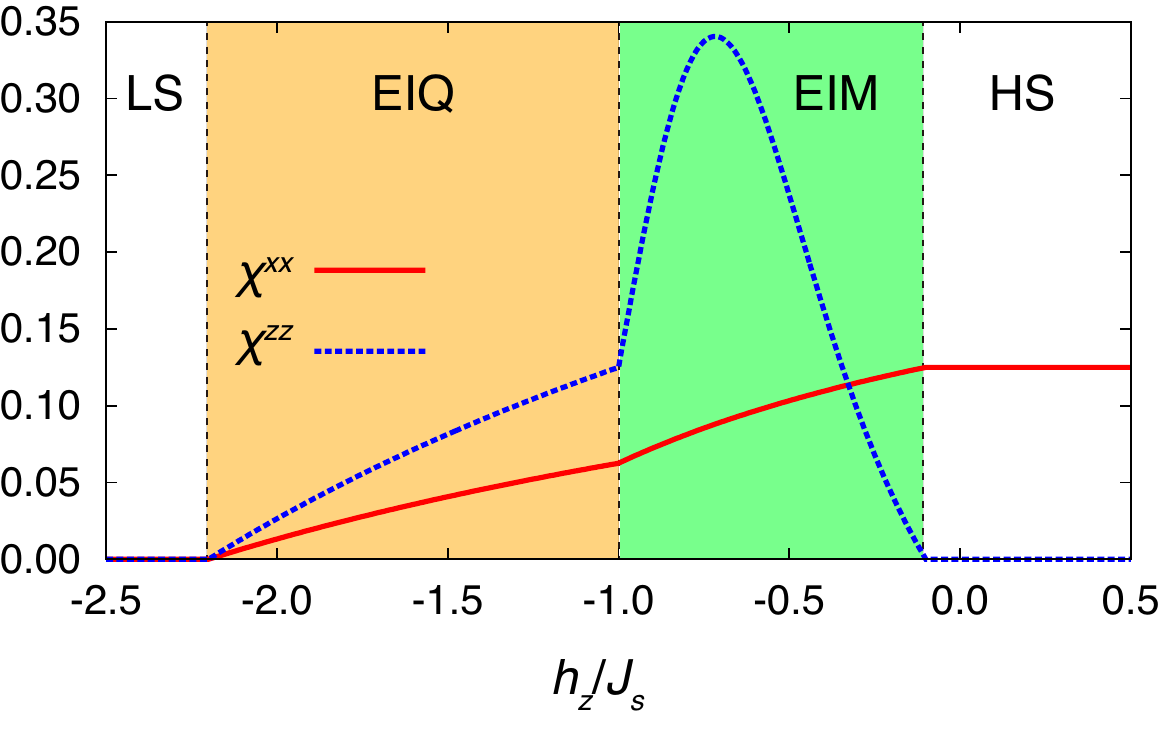}
 \caption{
 Transverse and longitudinal components of the static magnetic susceptibilities.
 The parameter values are chosen to be $J_z/J_s=0.1$, $J_x/J_s=0.5$ and $J_y=0$. 
The symbols HS, EIQ, and EIM indicate the AFM-HS phase, the EIQ phase, and the EIM phase, respectively.
}
\label{mf_sus}
\end{center}
\end{figure}

At the end of this subsection, we show the magnetic susceptibilities, which are derived from the dynamical spin correlation function given in Eq.~(\ref{eq:sus}). 
The transverse and longitudinal components of the susceptibilities, $\chi^{xx}$ and $\chi^{zz}$, are shown in Fig.~\ref{mf_sus}.
Vanishing of the susceptibilities in the LS phase originate from the spin gap.
In the AFM-HS phase, $\chi^{zz} =0$ and $\chi^{xx} \ne 0$ are attributed to the AFM order, in which the staggered moment is assumed to be along the $S^z$ axis. 
In the two EI phases, both $\chi^{xx}$ and $\chi^{zz}$ are finite. 
The finite magnetic susceptibilities are shown in the EIQ phase, corresponding to the Van Vleck components. 
The results are available to identify experimentally the EIQ phase in which the conventional magnetic orders do not emerge.  
Even in the AFM ordered state, $\chi^{zz}$ is finite in the EIM phase in contrast to the AFM-HS phase, since the longitudinal responses occur by changing a relative weight of the LS and HS states.  

\subsection{Goldstone modes}

In this subsection, we discuss the number of the Goldstone modes in each phase.
By following the results in Refs.~\cite{NIELSEN1976445,PhysRevLett.108.251602}, the number of the Goldstone modes termed $n_{\rm NG}$ are governed by the number of the broken symmetries termed $n_{\rm BS}$. 
The equation $n_{\rm NG}=n_{\rm BS}$ is satisfied in the case that $\means{[{\cal O}_\xi, {\cal O}_{\xi'}]}=0$ for all pairs of $\xi$ and $\xi'$.  
The SO(3) symmetry for the spin rotation is only the continuous symmetry in the present effective Hamiltonian, and the generators are $S_{\rm tot}^x$, $S_{\rm tot}^y$, and $S_{\rm tot}^z$, all of which commute with the Hamiltonian given in Eq.~(\ref{eq:2}).
The conditions $\means{[S_{\rm tot}^\alpha, S_{\rm tot}^\beta]} \propto i \varepsilon_{\alpha \beta \gamma} \means{S_{\rm tot}^\gamma}=0$, where $\varepsilon_{\alpha \beta \gamma}$ is the Levi-Civita antisymmetric tensor, are satisfied in all ordered phases appearing in the MF phase diagrams shown in Fig.~\ref{mf}.
Thus, the criteria derived in Ref.~\cite{PhysRevLett.108.251602} are available in the present phases. 
It is also shown in this case that the dispersion relation in the long-wave length limit is linear, unless the parameter set is artificially fine-tuned~\cite{PhysRevLett.108.251602}. 

It is trivial that there are not any Goldstone modes in the LS phase in which the SO(3) symmetry is not broken. 
There are also no Goldstone modes in the LS/HS ordered phase, where magnetic orders are not assumed in the present MF calculations.  
In the EIQ phase, the two of the three generators change the ordered state. 
For example, the wave function $\kets{\psi_{\rm EIQ}}=C_1\kets{L}+C_2\kets{H_Z}$ with real numbers $C_1$ and $C_2$ is not invariant by the operations of $S_{\rm tot}^x$ and/or $S_{\rm tot}^y$.
That is, $n_{\rm NG}=2$.  
In the EIM phase with the AFM order, all of the generators change the ordered state. 
It is explicitly shown that the wave functions for the $A$ and $B$ sublattices are given by $\kets{\psi_{\rm EIM}}_A=D_1\kets{L}+D_2\kets{H_X}+D_3\kets{H_Y}$ and $\kets{\psi_{\rm EIM}}_B=D_1\kets{L}+D_2^*\kets{H_X}+D_3^*\kets{H_Y}$ with a real number $D_1$ and complex numbers $D_2$ and $D_3$ [see Eqs.~(\ref{eq:13}) and~(\ref{eq:14})] are not invariant under the operations of $S_{\rm tot}^x$, $S_{\rm tot}^y$ and $S_{\rm tot}^z$.
 Therefore, $n_{\rm NG}=3$.  
Finally, in the AFM-HS phase, the number of the broken symmetry is two, for example, $S_{\rm tot}^x$ and $S_{\rm tot}^y$ in the case that the staggered moments are parallel to the $S^z$ axis. 
Thus, $n_{\rm NG}=2$. 
The above discussion is consistent with the numerical results shown in Fig.~\ref{band}.
In all phases, the linear dispersion relations around the $\Gamma$ point are predicted from the results in Ref.~\cite{PhysRevLett.108.251602} except for the case with fine-tuning of the parameter sets. 
This is also confirmed by the numerical calculations shown in Sec.~\ref{sec:dis}.

The present analyses are extended to the case without the pair hopping, in which the U(1) symmetry exists. 
The $z$ component of the PS operator $\tau^z_{\rm tot}(\equiv \sum_i \tau^z_i)$ commutes with the  Hamiltonian.
We find that $n_{\rm NG}=3$ and 4 in the EIQ and AFM-EIM phases, respectively, and $n_{\rm NG}$ in other phases are the same with those in the case of $I \ne 0$.

\section{Discussion and summary}
\label{sec:discussion-summary}

We discuss the relationships to the previous theoretical studies in the EI phases. 
The two-orbital Hubbard model with the finite-energy difference was studied by using the variational cluster approximation~\cite{PhysRevB.85.165135}.
Note that the condition $t_a/t_b>0$ in Ref.~\cite{PhysRevB.85.165135} is transformed into $t_a/t_b<0$ by changing signs of $J_x$ and $J_y$ in the present effective Hamiltonian.
The EI phase stabilized by the Hund coupling, which is termed the excitonic spin density wave (SDW) phase, was found in between the LS phase and the AFM-HS phase.
This EI phase may correspond to the EIQ phase in the present study, in which the local magnetic moment does not appear but the time-reversal symmetry is broken.
The symmetry analyses given in the present study are applicable to the original two-orbital Hubbard model as follows. 
It is trivial that the time reversal and translational symmetries are retained. 
The $Z_2$ symmetry for the PS operators proposed in the present study corresponds to the transformations $c_{ia\sigma}^\dagger \to -c_{ia\sigma}^\dagger$ and $c_{ia\sigma} \to -c_{ia\sigma}$ for all $i$ and $\sigma$ in the two-orbital Hubbard model. 
The first-order phase transition between the excitonic SDW and AFM-HS phases, and the second-order transition between the excitonic SDW and the LS phases shown in ~\cite{PhysRevB.85.165135} are understood in the symmetry relations clarified in the present study. 
No phases corresponding to the EIM phase associated with the AFM order as well as the LS/HS ordered phase in the present study were shown in Ref.~\cite{PhysRevB.85.165135}.
It is shown in Fig.~\ref{mf} that the difference between the band widths is necessary to realize these two phases.
Another EI phase, in which the time-reversal symmetry is not broken, termed the excitonic CDW phase was studied in Refs.~\cite{RevModPhys.40.755,PhysRevB.90.245144}.
This phase is understood in the present framework as a phase where the LS and a spin-singlet state given by 
\begin{align}
\kets{S}=\frac{1}{\sqrt{2}}\left(c_{a\uparrow}^\dagger c_{b\downarrow}^\dagger - c_{a\downarrow}^\dagger c_{b\uparrow}^\dagger\right)\kets{0},
\label{eq:singlet}
\end{align} 
are mixed in the wave function.
Since the energy of this singlet state is much higher than $\kets{L}$ and $\kets{H_\Gamma}$ introduced in Eq.~(\ref{eq:hs-1}) at vicinity of the LS-HS phase boundary, this state is irrelevant to the low-energy electronic structure in the present model.
This is reasonable that the EI phases with the time-reversal symmetry breaking are favored in the case that the Hund coupling is comparable to the crystalline field splitting~\cite{PhysRevB.90.245144}. 
On the other hand, the excitonic CDW phase may be stabilized by the Jahn-Teller type electron-lattice interaction, which is not included in the present theory.

\begin{figure}[t]
\begin{center}
\includegraphics[width=\columnwidth,clip]{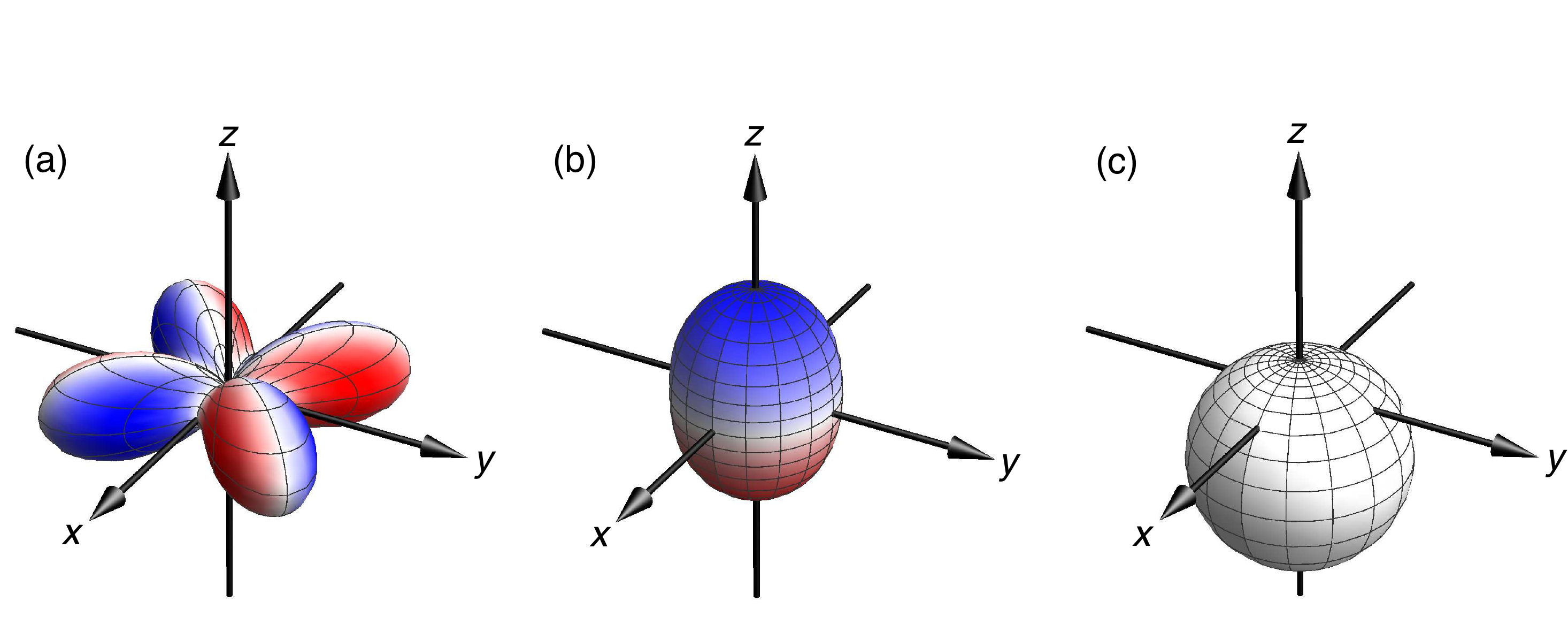}
 \caption{
 Charge and spin distributions at a single site in the EI phases. 
 In (a) and (b), the wave functions are given as $\frac{1}{\sqrt{2}}\left(\kets{L}+\kets{H_{Z}}\right)$, where the $a$ and $b$ orbitals are taken to be (a) the $d_{x^2-y^2}$ and $d_{xy}$ orbitals, and (b) the $p_{z}$ and $s$ orbitals, respectively. 
In (c), the wave function is given as $\frac{1}{\sqrt{2}}\left(\kets{L}+\kets{S}\right)$, where the $a$ and $b$ orbitals are taken to the $p_{z}$ and $s$ orbitals, respectively. 
Radius and color represent the charge and spin distributions, respectively.
Red and blue regions represent positive and negative $S^z$, respectively.
}
\label{orbital}
\end{center}
\end{figure}

From the orbital physics viewpoint, the excitonic phases are characterized in terms of the multipole order.
Let us consider some examples where the explicit symmetries of the $a$ and $b$ orbitals are assumed. 
First, we consider the case where the $a$ and $b$ orbitals are the $d_{x^2-y^2}$ and $d_{xy}$ orbitals, respectively, which have the same parity in the wave functions. 
The charge and spin distributions in the EIQ phase are shown in Fig.~\ref{orbital}(a), 
where the wave function is given by $\frac{1}{\sqrt{2}}\left(\kets{L}+\kets{H_{Z}}\right)$ with  $\kets{L}\sim c_{b\uparrow}^\dagger c_{b\downarrow}^\dagger\kets{0}$. 
In this figure, the radius and color represent the charge distribution and the spin distribution for the $S^z$ component, respectively. 
The charge distribution shows a similar symmetry to the $d_{xy}$ orbital. 
On the other hand, there are eight nodes in the spin distribution, indicating that the magnetic triakontadipole ($2^5$-pole) moment as a manifestation of the time-reversal symmetry breaking.
In Fig.~\ref{orbital}(b), we present the charge and spin distributions in the case where the $a$ and $b$ orbitals are chosen as the $p_{z}$ and $s$  orbitals, respectively, which have the odd and even parities.
The EIQ phase given by $\frac{1}{\sqrt{2}}\left(\kets{L}+\kets{H_{Z}}\right)$ is considered. 
There appears a finite magnetic dipole moment, but no electric dipole moment.
% The charge distribution shows the even parity and no electric dipole moment appears. 
Characteristics in the charge distribution are different in the EIQ phase where the spin singlet state introduced in Eq.~(\ref{eq:singlet}) is included.
In Fig.~\ref{orbital}(c), we present the charge and spin distributions in the EI phase where the $a$ and $b$ orbitals are chosen to be the $p_{z}$ and $s$ orbitals, respectively, and the wave function is given by $\frac{1}{\sqrt{2}}\left(\kets{L}+\kets{S}\right)$, which might correspond to the excitonic CDW phase.
The spin dipole moment does not appear but the electric dipole moment does.  
This might correspond to the ferroelectric state discussed in the Falicov-Kimball model in Ref.~\cite{PhysRevLett.89.166403}.

The present theoretical studies for the EI phases have direct implications of the candidate materials of the EI state.
The most relevant material is a series of the perovskite cobalt oxides, $R_{1-x}A_x$CoO$_3$ and their family, in which the localized orbital picture works well. 
The low temperature insulating phase in Pr$_{0.5}$Ca$_{0.5}$CoO$_3$, where any experimental indications for the magnetic and charge orders are not observed until now, is one of the possible candidates of EI phases~\cite{doi:10.1143/JPSJ.73.1987,hejtmanek2013phase,PhysRevB.89.115134,PhysRevB.90.235112}.
A possibility of the EI phase should be also reexamined in LaCoO$_3$. 
The ground state is confirmed experimentally as a non-magnetic LS band insulator. 
We expect that LaCoO$_3$ is located near the phase boundary between the LS and EIQ phases in Fig.~\ref{mf}.
This is because the NN Co-O bond length, which gives an indication of the spin-state character, 
is almost the same value with that in Pr$_{0.5}$Ca$_{0.5}$CoO$_3$, a candidate material of EI~\cite{PhysRevB.66.052418}.
This fact implies that the LS state in LaCoO$_3$ is possibly changed into the EIQ phase by changing an energy balance between the Hund coupling and the crystalline field splitting.
Applying the magnetic field and/or expanding the lattice constant by chemical substitution or by utilizing the thin-film technique on a substrate are the plausible routes. 
Recently, a new phase is found under the strong high-magnetic field in LaCoO$_3$~\cite{doi:10.1143/JPSJ.78.093702,Ikeda2015pre}. 
One possible scenario of this phase is a magnetic-field induced EI phase, in which the Zeeman energy is gained due to the HS component mixed with the LS state.
On the other hand, there are some limitations in the present study based on the two-orbital Hubbard model, where the realistic five orbitals are simplified to the two orbitals. 
Detailed comparison with theoretical calculations are necessary to confirm % this
the above scenario.

The present orbital physics viewpoints for the EI system are applicable not only to cobaltites~\cite{PhysRevB.89.115134,PhysRevB.90.235112,chaloupka2008unusual} but also to other candidate materials, such as layered chalcogenides~\cite{PhysRevLett.103.026402,Wakisaka2012,PhysRevB.87.035121,PhysRevB.90.245144,PhysRevLett.105.176401,PhysRevB.91.205135,porer2014non}, iron-based superconductors~\cite{0295-5075-86-1-17006,doi:10.1143/JPSJS.77SC.158,PhysRevB.79.180504,PhysRevLett.110.207205}, dimer-type organic compounds~\cite{doi:10.1143/JPSJ.77.074709,PhysRevB.91.245132,yamakawa2016novel,PhysRevB.82.125119,lunkenheimer2012multiferroicity}, and others.
In some of these materials, the valence and conduction bands originate from atomic orbitals in different lattice sites.
The charge and spin multipoles in this case are represented by generalization of the PS operators introduced in Sec.~\ref{sec:model}, where the electronic operators are replaced by $c_{ia}^\dagger$ and $c_{j b}^\dagger$ with different sites $i$ and $j$ together with a structure factor.
The electronic ferroelectriciy proposed in the dimer-type organic salts is understood from the orbital physics concept for the EI system.
In low-dimensional organic molecular salts, such as series of Tetramethyl-tetrathiafulvalene (TMTTF)~\cite{PhysRevLett.85.1698,PhysRevLett.86.4080,doi:10.1143/JPSJ.76.103701,doi:10.1143/JPSJ.77.113705} and Bis(ethylenedithio)tetrathiafulvalene (BEDT-TTF) compounds~\cite{doi:10.1143/JPSJ.77.074709,PhysRevB.91.245132,yamakawa2016novel}, the molecular dimer units builds a framework of a crystal lattice.
One electron per dimer occupies a bonding molecular orbital in each dimer unit.
Since the Coulomb interaction between electrons inside a dimer unit is larger than the band width, the system is classified as a Mott insulator, which is termed a dimer-Mott insulator.
Recently, the experimentally observed dielectric anomalies are considered to be attributable to the electric dipole moment insider of the dimer-units, in which the electronic charge distributions are polarized cooperatively~\cite{PhysRevB.82.125119,lunkenheimer2012multiferroicity,PhysRevB.82.241104,doi:10.1143/JPSJ.79.063707}.
From the viewpoint of the EI phase, this ferroelectric phase is realized by a spontaneous mixing of the bonding and antibonding orbital bands.
The collective excitation modes and the superconductivity due to the charge fluctuation studied so far~\cite{PhysRevLett.110.106401,PhysRevB.87.085133,doi:10.7566/JPSJ.84.023703} should be reexamined as an EI system.

The present study provides ways to identify the EI phases. 
The experimental observations of the EI order parameters, i.e., the electric and/or magnetic multipoles, using the resonant x-ray scattering method, are the direct evidence of the EI phase, although the detailed polarization analyses are necessary. 
The EI transition is classified as the Ising-like transition due to the breaking of the $Z_2$ symmetry. 
Thus, above the transition temperature, the dominant excitations are the diffusive modes which may appear in the quasi-elastic component in the inelastic x-ray scattering measurements. 
The soft-excitation modes above the transition temperature are possible to appear through the coupling between the electronic orbital and the lattice distortion. 
This is similar to the PS-phonon coupled system in the order-disorder type ferroelectricity~\cite{doi:10.1143/JPSJ.36.641}.

The observations of the collective excitations and their responses by the external stimuli also distinguish the EI phase from the conventional band and Mott insulators. 
Whereas the conventional magnetic order does not appear in EIQ phase, the magnetic gapless modes exist. 
These modes contribute to the magnetic susceptibility and some thermodynamic quantities, and are directly detected by the inelastic neutron/magnetic x-ray  scatterings. 
One remarkable difference from the conventional magnetic ordered phase is seen in the longitudinal-magnetic excitation modes; 
these modes are active in the EI modes, in contrast to the AFM-HS phase as shown in Fig.~\ref{spec}.
The longitudinal modes are attributable to changes of the relative weights of the LS and HS states in the wave function. 

In summary, we have investigated the ground-state phase diagram and the collective excitations in the EI system from the orbital physics viewpoint in strongly correlated electron system. 
The effective Hamiltonian for the low-energy electronic structure is derived from the two-orbital Hubbard model with the finite-energy difference between the orbitals. 
This is represented by the spin operators of $S=1$ and the PS operators for the spin-state degrees of freedom. 
The two kinds of the EI phases, termed the EIQ and EIM phases, are realized as a consequence of the competition between the crystalline field effect and the Hund coupling, in addition to the LS band insulating phase, the AFM-HS Mott insulating phase and the LS/HS ordered phases.
The EI phase transition is identified as an Ising-like transition due to the spontaneous breaking of the $Z_2$ symmetry. 
The magnetic structures are distinguished in the two EI phases; 
the AFM order in the EIM phase and the spin-nematic order in the EIQ phase. 
The Van Vleck component in the magnetic susceptibility appears in the EI phases. 
The longitudinal magnetic modes as well as the transverse modes are active in the two EI phases, in contrast to the LS and AFM-HS phases. 
These characteristics are available to identify the EI phases using the inelastic neutron and magnetic x-ray scatterings.
The orbital physics viewpoints proposed in the present paper are applicable to a wide class of the candidate materials of the EI systems. 

 \begin{acknowledgments}
The authors would like to thank Y. Ohta, H. Fukuyama, H. Sawa, M. Sato, T. Momoi, and J. Ostuki for fruitful discussions. 
This material is supported by Grant-in-Aid for Scientific Research (No. 26287070) from MEXT. 
A part of the numerical calculations has been performed using the supercomputing facilities at ISSP, the University of Tokyo. 
 \end{acknowledgments}

\appendix

\section{Exchange constants in the effective Hamiltonian}
\label{sec:exchange-parameters}

\begin{figure}[t]
\begin{center}
\includegraphics[width=\columnwidth,clip]{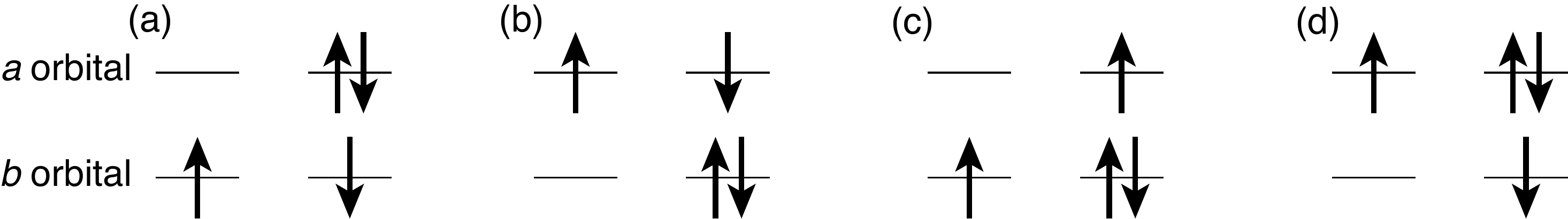}
 \caption{
Schematic spin and orbital configurations in the intermediate states in the perturbation processes for (a) $\kets{\psi_1}$, (b) $\kets{\psi_2}$, (c) $\kets{\psi_3}$, and (d) $\kets{\psi_4}$. 
Arrows represents directions of the electronic spins. 
}
\label{excited}
\end{center}
\end{figure}

In this appendix, we present the explicit forms of the exchange constants in the effective Hamiltonian in Eq.~(\ref{eq:2}).
This Hamiltonian is derived by the second order perturbation processes in terms of ${\cal H}_t$ in Eq.~(\ref{eq:exH}). 
The LS and HS states given by Eqs.~(\ref{eq:ls})--(\ref{eq:hs-1}) are the bases of the effective Hamiltonian. 
The energies of the LS and HS states are denoted by $E_L(=U+\Delta-\Delta')$ and $E_H(=U'+\Delta'-J)$, respectively. 
The four electron configurations termed $|\psi_n \rangle$ $(n=1-4)$
shown in Fig.~\ref{excited} are adopted as the intermediate states of the perturbation processes. 
The corresponding energies of the intermediate states are denoted as 
$E_1=E_2=2\Delta+2U'+U-J$, 
$E_3=\Delta+2U'+U-J$, and $E_4=3\Delta+2U'+U-J$ .  
Then, the exchange constants in the effective Hamiltonian are given by 
\begin{align}
E_{0}&=\frac{N(E_{L}+E_{H})}{2}+\frac{zN(2\delta E_{LH}-9\delta E_{LL}-9J_s)}{32},\\
h_z&=\frac{E_{L}-E_{H}}{4}-\frac{z(2\delta E_{LH}+\delta E_{LL}-3J_s)}{16},\\
J_z&=\frac{2\delta E_{LH}-\delta E_{LL}-J_s}{16},\\ 
J_s&=\frac{t_a^2+t_b^2}{U+J},\\
J_x&=-\frac{J'+J''}{2},\\
J_y&=\frac{J'-J''}{2} .
\end{align}
Here, we define 
%\begin{widetext}
\begin{align}
\delta E_{LL}&=\frac{4f^2g^2(t_a^2+t_b^2)}{2U'-U-J+2\Delta'}, 
\label{eq:dELL} \\
 \delta E_{LH}&=
(t_a^2+t_b^2)\left[\frac{f^2}{-\Delta+U'+\Delta'}+\frac{g^2}{\Delta+U'+\Delta'}\right], 
\label{eq:dELH}\\
J'&=2t_a t_b fg\left[\frac{1}{U+J}+\frac{1}{2U'-U-J+2\Delta'}\right],
\label{eq:10}\\
J''&=t_a t_b\left[\frac{f^2}{-\Delta+U'+\Delta'}+\frac{g^2}{\Delta+U'+\Delta'}\right] .
\label{eq:11}
\end{align}
%\end{widetext}
Constants $f$ and $g$ are introduced in Eq.~(\ref{eq:ls}).
We note that $J'$ and $J''$ given in Eq.~(\ref{eq:10}) and Eq.~(\ref{eq:11}), respectively, are proportional to $t_at_b$ and have the same sign with each other, i.e., $J'J''>0$.
As a consequence, we have a relation $|J_x|\geq|J_y|$. 
The equal sign, $J_x=J_y$, is satisfied in the case of $I=0$ which leads to $g=0$ and $J'=0$.

We introduce the physical meanings of the energy constants given in Eqs.~(\ref{eq:dELL})-(\ref{eq:11}). 
In the MF approximation introduced in Sec.~\ref{sec:method}, 
the energies in the LS, AFM-HS, and LS/HS ordered phases are given by 
\begin{align}
 E_L^{\rm MF}&=N E_L-\frac{zN\delta E_{LL}}{2}, \\
 E_H^{\rm MF}&=NE_H-J_s zN, \\
 E_{LH}^{\rm MF}&=\frac{N}{2}(E_H+E_L)-\frac{zN\delta E_{LH}}{2}, 
\end{align}
respectively. 
Therefore, $\delta E_{LL}$ and $\delta E_{LH}$ represent corrections from the MF energies in the LS and LS/HS ordered phases due to the second order perturbational processes.

The last two terms in the effective Hamiltonian in Eq.~(\ref{eq:9}) are also represented by using the projection operators $\bm{d}^\dagger=(d_X^\dagger, d_Y^\dagger, d_Z^\dagger)$ introduced in Sec.~\ref{sec:model} as
\begin{align}
J'\sum_{\langle ij \rangle} \left ( {\bm d}_i^\dagger \cdot {\bm d}_j ^\dagger +{\rm H.c.} \right )
+J''\sum_{\langle ij \rangle}\left (  {\bm d}_i^\dagger {\bm d}_j  +{\rm H.c.} \right ) . 
\end{align}
The first term represents changes in the spin states in the NN sites, where the LS states are changed into the HS states in both the $i$ and $j$ sites and vice versa.
The second term represents the exchange of the LS and HS states between the NN sites.  

\bibliography{refs}

\end{document}